\documentclass[11pt,a4paper]{article}
\usepackage{amsmath,amsfonts,amssymb,graphicx,setspace}
\usepackage{hyperref}
\usepackage[top=1.1in,bottom =1.1in,left=1.1in,right=1.1in]{geometry}

\usepackage{grffile}
\usepackage{latexsym}
\usepackage{color}
\usepackage{amsbsy}
\usepackage{authblk}
\usepackage{epstopdf}
\usepackage{cite}
\usepackage{mathtools}

\def\au{ {\textbf{u}}}
\def\aut{ {\textbf{z}}}
\def\ax{ {\textbf{x}}}
\def\ay{ {\textbf{y}}}
\def\av{ {\textbf{v}}}
\def\aw{ {\textbf{w}}}

\def\ao{ {\textbf{0}}}

\def\bigS{ {\mathcal {S}}}

\newcommand{\R}{\mathbb R}

\def\be#1\ee{\begin{equation}#1\end{equation}}
\newcommand{\fer}[1]{(\ref{#1})}
\newtheorem{theorem}{Theorem}[section]

\newtheorem{remark}[theorem]{Remark}

\newenvironment{equations}{\equation\aligned}{\endaligned\endequation}
\def\bqa{\begin{eqnarray}}
\def\eqa{\end{eqnarray}}

\def\a{\alpha}

\def\b{\beta}

\def\va{\varphi}
\def\ab{{\alpha,\beta}}
\def\ab1{{\alpha_1,\beta_1}}

\newcommand{\bd}{\begin{displaymath}}
\newcommand{\ed}{\end{displaymath}}
\newcommand{\ba}{\begin{eqnarray}}
\newcommand{\ea}{\end{eqnarray}}

\def\R{\mathbb{R}}

\newcommand\rev[1]{\textcolor{black}{#1}}

\title{Modelling contagious viral dynamics: a kinetic approach based on mutual utility}

\author{Giulia Bertaglia\thanks{Department of Environmental and Prevention Sciences, University of Ferrara, Italy. \par e-mail: \texttt{giulia.bertaglia@unife.it}} \quad
Lorenzo Pareschi\thanks{Maxwell Institute and Department of Mathematics, Heriot-Watt University, Edinburgh, UK. \par e-mail: \texttt{l.pareschi@hw.ac.uk}} \thanks{Department of Mathematics and Computer Science, University of Ferrara, Italy. \par e-mail: \texttt{lorenzo.pareschi@unife.it}} \quad
Giuseppe Toscani\thanks{Department of Mathematics, University of Pavia, Italy.  \par e-mail: \texttt{giuseppe.toscani@unipv.it} } \thanks{IMATI, Institute for Applied Mathematics and Information Technologies ``Enrico Magenes'', Pavia, Italy}}

\date{\today}

\begin{document}
\maketitle

\begin{center}\small
\parbox{0.85\textwidth}{

\textbf{Abstract}. 
The temporal evolution of a contagious viral disease is modelled as the dynamic progression of different classes of population \rev{with individuals interacting pairwise. This interaction follows a binary mechanism typical of kinetic theory, wherein agents aim to improve their condition with respect to a mutual utility target}. To this end, we introduce kinetic equations of Boltzmann-type to describe the time evolution of the probability distributions of the multi-agent system. \rev{The interactions between agents are defined} using principles from price theory, specifically employing Cobb-Douglas utility functions for binary exchange and the Edgeworth box to depict the common exchange area where utility increases for both agents.
\rev{Several numerical experiments presented in the paper highlight the significance of this mechanism in driving the phenomenon toward endemicity.}

\medskip

}
\end{center}

\tableofcontents

\section{Introduction} 
Complex processes in biological systems at microscopic and macroscopic scales can be fruitfully described by means of methods of kinetic theory^^>\cite{Ari}. \rev{Indeed, while kinetic theory was initially introduced to describe the collisional dynamics of rarefied gas molecules, in recent years, it has been successfully applied in various research fields. This application aims to model the interactions involved in the dynamics of multi-agent systems composed of a large number of individuals, allowing the study of emergent collective phenomena and self-organization patterns^^>\cite{DPTZ,DTZ,FPZ,PTbook}.}

Furthermore, as recently discussed in^^>\cite{Ari2}, in many cases the concept of entropy plays an important role \rev{in biological processes}. In addition to examples related to irreversible chemical reactions (where irreversibility can be described in terms of the monotonicity of entropy^^>\cite{Ziv}), other concepts, like measures of information, are frequently discussed in relation to living systems and the influence on fluctuations (cf.^^>\cite{Dem} and references therein).

In this paper, we address irreversibility in biological processes \rev{(referring here to phenomena in which certain changes within the system cannot be reversed)} through the concept of \emph{utility}, \rev{proposing an approach that draws inspiration} from a well-known model in microeconomics: the Edgeworth box^^>\cite{TBD}.

Although the modeling introduced here is adaptable for describing various biological phenomena with irreversible nature, in this manuscript, we will focus on viral evolution, \rev{where irreversibility is inherent in the contagious dynamics itself.} This choice is motivated by the recent interest in infectious dynamics, particularly heightened in the wake of the COVID-19 pandemic. \rev{See, for instance, references \cite{Bert4,Bert5,Bert}, where different methodologies are proposed to study the spread of an infectious disease through multiscale modeling, also considering uncertainties in parameters; \cite{Eli}, where the spread of COVID-19 is studied through delay differential equations; and \cite{TAKT}, where a "nesting modeling" approach is proposed}. Although the evolution of Covid-19, as in the case of other diseases with seasonal characteristics, \rev{such as influenza, measles, pertussis, mumps, diphtheria, varicella, and scarlet fever^^>\cite{Low,Met}}, did not manifest a monotonic behaviour in moving from a high-risk pandemic phenomenon to a low-risk endemic phenomenon, but rather a behaviour similar to that of damped oscillations (successive pandemic waves with decreasing impact^^>\cite{Vig}), the irreversibility of the phenomenon remains evident.
\rev{The methodology here adopted} is complementary to others, such as to the one proposed in ^^>\cite{Bohl, Cas}, where game theory is used to model the viral behaviour, \rev{or the modelling framework presented in^^>\cite{DLT23,DLT,LoyTosin} to track the viral load evolution through kinetic equations for multi-agent systems}.



In this work, we propose to study the evolution of an epidemic \rev{through multi-agent modeling. In this framework, healthy and infected individuals (i.e., agents) interact with each other, determining the spread of the virus (for an introduction to kinetic models in epidemic dynamics, refer to \cite{Esurvey}). Within this modeling framework typical of kinetic theory, we introduce a new dynamic of ``utility" that governs the interaction between the agent and the virus itself. We assume that this interaction is driven by the mutual benefit for both the individual and the virus to improve their own socio-physical/viral condition.} For the individual, this involves finding an optimal balance between sociality and the risk of contracting the disease, while for the virus, it entails maintaining an optimal balance between contagiousness and minimizing negative effects on the host.

To this aim, we introduce kinetic equations of Boltzmann type designed to describe the evolution in time of the probability distributions of the agent-virus system undergoing \emph{binary interactions}. The leading idea is to describe the interactions by means of some fundamental rules in price theory, in particular by using Cobb-Douglas utility functions for the binary exchange, and the Edgeworth box for the description of the common exchange area in which utility is increasing for both agents \rev{(further details on both Cobb-Douglas utility function and Edgeworth box are given in Section \ref{Edge})}.

\rev{As a side result of our modelling, we will introduce a way to derive classical compartmental models directly from a Boltzmann-type equation.} This allows us to assess the irreversibility of the disease resulting from the ongoing interactions, by referring to classical \rev{SI and SIR models} of compartmental epidemiology, \rev{two statistical models dating back} to the pioneering work of Kermark and McKendrick^^>\cite{KM}. However, despite an increase in simulation complexity, the proposed approach can be easily extended to other compartmentalizations, better suited for studying specific real epidemic phenomena^^>\cite{Bert3,Bert2}. \rev{In this paper, we will limit ourselves to the extension that considers the loss of immunity in recovered individuals and the additional presence of the compartment of deceased people}. Numerical experiments will help clarify the importance of these mechanisms in driving the phenomenon towards endemicity.

The rest of the manuscript is organized as follows. In Section 2 we describe the general structure of the mathematical model under consideration and illustrate its relationships with classical compartmental models. Then, Section 3 is devoted to present the competitive mechanism at the basis of the viral evolution inspired by the Edgeworth box theory in economics. We illustrate the behaviour of the models with the aid of some numerical experiments in Section 4. Some final remarks conclude the manuscript.

\section{The model}\label{model}

\rev{
In this section, we will first present the modelling framework to study the temporal evolution of a viral disease in the context of the Boltzmann-type equation of kinetic theory \cite{PTbook}. The presentation will be made in a very general context to emphasize the broad scope of the proposed modelling, that can be adapted to the description and study of numerous irreversible biological processes. Next, we will demonstrate how it is possible to derive a compartmentalization approach, typical of the epidemiological literature, from the proposed model. This will be done first in the case of an SI subdivision of the population and then for the classic SIR compartmentalization.}

\begin{figure}[t]
\centering
\includegraphics[width=0.9\textwidth]{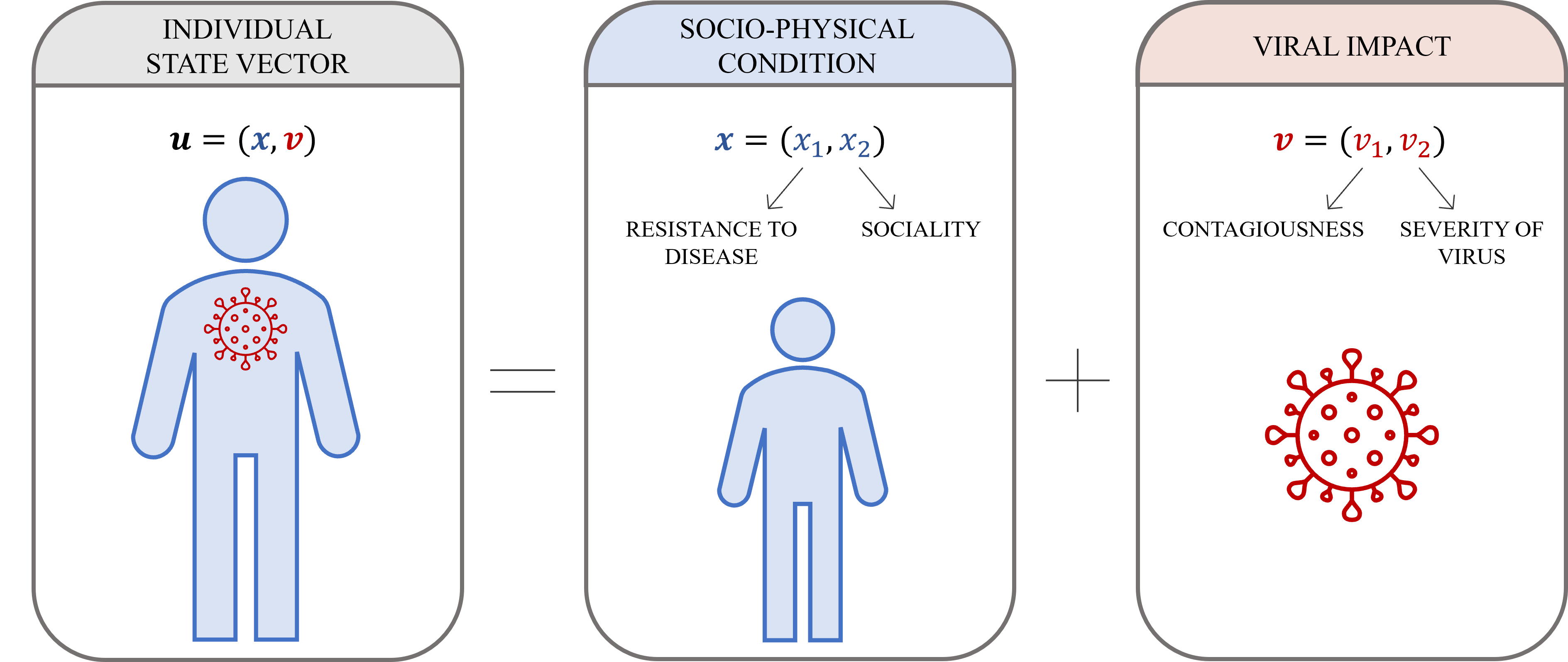}
\caption{Schematic summary of the components of the individual state vector. The individual state is identified by a joint state vector $\au =(\ax,\av)$, where the \emph{socio-physical condition} vector $\ax =(x_1,x_2)$, dependent on the personal rate of resistance to the disease ($x_1$) and the degree of sociality ($x_2$), refers to the individual, while the \emph{viral impact} vector $\av =(v_1,v_2)$, composed by the degree of contagiousness ($v_1$) and the severity of the disease ($v_2$), refers to the virus.}
\label{fig:SV}
\end{figure}

\subsection{A Boltzmann-type dynamics}
Given a population of individuals, we denote by $f(\au, t)$, with $\au = (u_1, u_2\dots,u_n)$, $n\ge 2$,  the statistical distribution of individuals which are characterized by a state vector $\au\in \R^n$ at time $t \ge 0$. 

Having in mind to study the evolution of an epidemic, and believing it important to be able to characterize its evolution in terms of both its infectiousness and its severity, we identify the state vector as the joint state vector $\au =(\ax,\av)$, where the two-states vectors $\ax =(x_1,x_2)$ and $\av =(v_1,v_2)$ refer to the individual and to the virus, respectively. 

In detail, we characterize the state of individuals in terms of their mental and physical well-being. To this aim, we indicate with $x_1 \ge 0$ the personal rate of resistance to disease, which generally depends on both physical state and age, and with $x_2 \ge 0$ the individual's degree of sociality, which can be measured by the mean number of social contacts during a fixed period of time. We will give this state $\ax$ the name of \emph{socio-physical condition}.

Likewise, following^^>\cite{Morens}, we characterize the state of the virus in terms of its main features, represented by  the degree of infectiousness, measured by $v_1 \ge 0$, and  the degree of severity of the disease, measured by $v_2 \ge 0$.  We will give this state $\av$ the name  of \emph{viral impact}. Clearly, individuals that have not been infected will be characterized by a viral impact $\av = \ao$.

A schematic summary of the components of the state vector is presented in Figure \ref{fig:SV}.
In what follows we will assume that all the state variables can vary between 0 and 1.

\begin{figure}[t]
\centering
\includegraphics[width=0.6\textwidth]{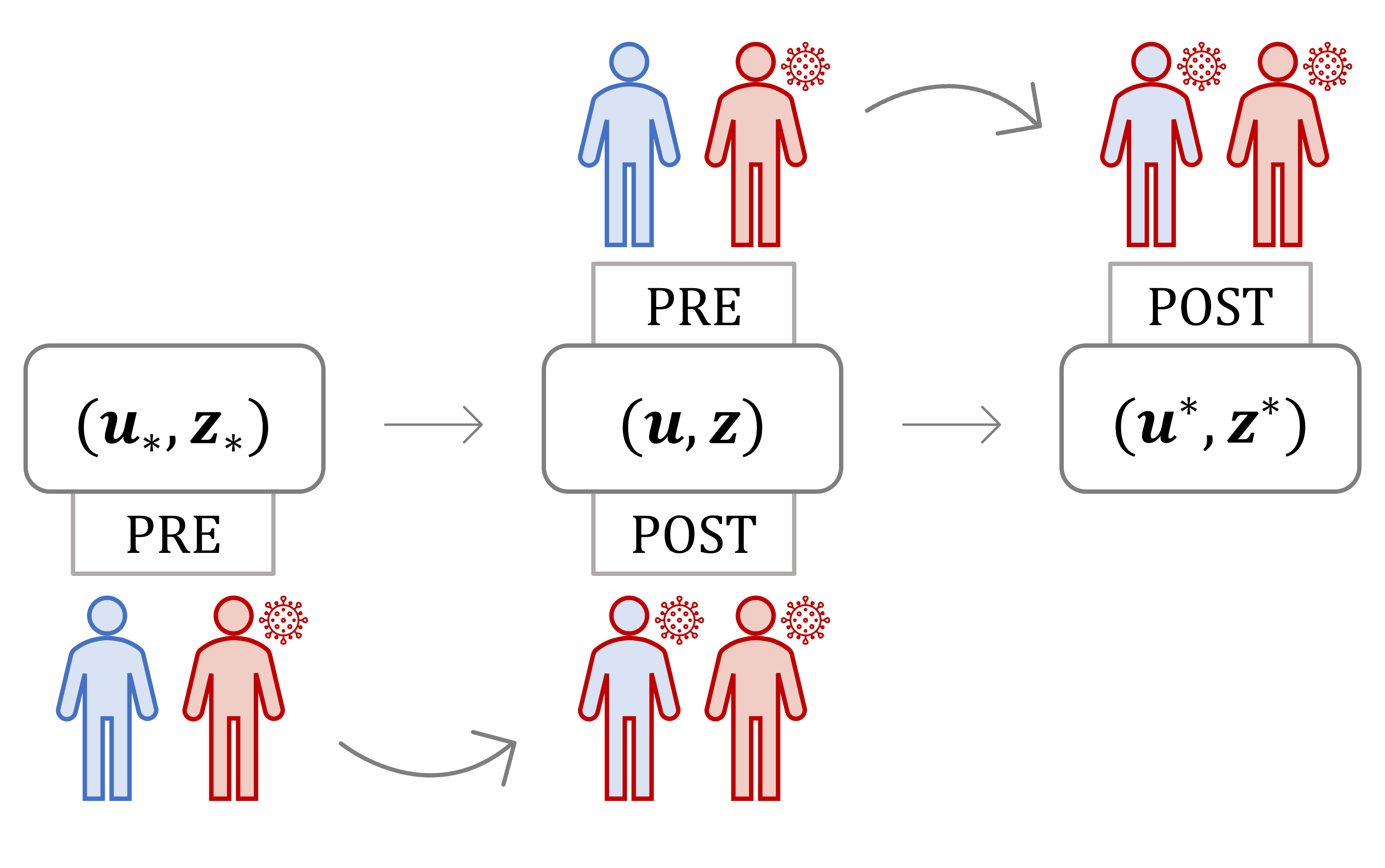}
\caption{\rev{Diagram of the interaction process between two agents. If the agents' pre-interaction states are represented by the pair $(\au_*,\aut_*)$, their interaction produces the post-interaction states $(\au,\aut)$. If we now take the pair $(\au,\aut)$ as the starting point pre-interaction, the dynamics leads the states to the post-interaction values $(\au^*,\aut^*)$.}}
\label{fig:transform}
\end{figure}

The evolution of the density $f$, due to the interaction of two agents characterized by the initial \rev{state vectors $\au =(\ax,\av)$ and $\aut =(\ay,\aw)$, respectively}, can be described by a homogeneous Boltzmann-type dynamics \rev{(please refer to^^>\cite{Cer,PTbook} for an introduction on kinetic theory and Boltzmann-type equations)}:
\rev{
\be
\label{Boltzmann}
\frac{\partial f(\au,t)}{\partial t} = Q(f,f)(\au,t),
\ee}
where $Q(f,f)$ is the bilinear interaction/collisional operator, defined as
\rev{
\be
\label{collision}
Q(f,f)(\au,t) = \int_{\R_+^4}\left[ \kappa_*(\au_*,\aut_*)f(\au_*,t)f(\aut_*,t) - \kappa(\au,\aut) f(\au,t) f(\aut,t)\right] \,d\aut .
\ee
Here, the starred state vectors $\au_*$ and $\aut_*$ are the pre-interaction states of the two agents, and $\kappa(\cdot,\cdot)$ is an appropriate function defining the kernel of the interaction, which may depend, in general, by all the state variables of the agents (see Figure \ref{fig:transform}).}

\rev{In \fer{collision}, $\kappa_*$ is related to $\kappa$ through the irreversible transformation $(\au_*,\aut_*) \to (\au,\aut)$, and it is defined as
\be\label{k*}
\kappa_*(\au_*,\aut_*) \coloneqq \kappa(\au_*,\aut_*) \frac 1{\mathcal{J}(\au_*,\aut_*)},
\ee
where $\mathcal{J}$ is the Jacobian of the \emph{inverse} transformation from the non-starred variables to the starred ones, so that
\be\label{jacobian}
d\au\,d\aut = \mathcal{J}(\au_*,\aut_*)\,d\au_*\, d\aut_*.
\ee}

As typical in kinetic theory^^>\cite{PTbook}, the interaction operator can be seen as the difference between a gain and a loss term of the evolutionary dynamics: 
\[Q(f,f) = Q^+(f,f) - Q^-(f,f) .\]
In this case, the two operators read
\rev{
\begin{equations}
\label{gain-loss}
Q^+(f,f)(\au,t) &= \int_{\R_+^4} \kappa_*(\au_*,\aut_*)f(\au_*,t)f(\aut_*,t)\, d\aut\\
Q^-(f,f)(\au,t) &=  f(\au,t) \int_{\R_+^4} \kappa(\au,\aut)  f(\aut,t) \,d\aut .
\end{equations}
}

We conclude this brief introduction to the model \rev{anticipating} that, in the infectious dynamics \rev{we will consider in this work}, the operator \eqref{collision} \rev{will be} non-zero only when an infected and a not yet infected individual (susceptible to infection) are interacting with each other. In all the other cases, the interaction dynamics \rev{will} not produce any effective update of the state of the agents. 

\rev{
\begin{remark}
We emphasize that the modelling framework presented so far describes an entirely generic phenomenon of binary interaction between agents. The characterization of the specific dynamics, in fact, requires the definition of the transformation laws 
\be
\au_*=\Phi(\au,\aut),\qquad \aut_*=\Psi(\au,\aut)
\label{eq:bin}
\ee
and their inverse, which remain undefined for the moment, but will be discussed in depth in the next sections.
\end{remark}
}
\subsection{Characterization as compartmental models}
\label{sec:compartments}
We now characterize the modelling just described by considering a classical partitioning of the population between susceptible and infected (\rev{SI}), which allows us to highlight how the standard compartmental approach widely used in the mathematical epidemiological literature can be derived from the proposed model.


In such a compartmental model, the initial population is divided into susceptible (S), who can contract the disease, and infected (I), who have already contracted it and can transmit it.
In terms of state variables of the kinetic approach introduced in the previous section, for a given time $t \ge 0$, we will assume that the distribution $f_S$ of the population of susceptible individuals depends only on their socio-physical conditions, as given by the state vector $(\ax,\ao)$. To maintain memory that distributions are generally dependent on the full vector $\au$, we will write 
\rev{
\be\label{fS}
f_S =f_S(\au, t)=f_S(\ax, \av, t) = f(\ax,\av,t)\cdot \delta(\av = \ao),
\ee
}
where $\delta(\av = \ao)$ denotes the Dirac delta function located in the point $\av =\ao$.  Unlike the class of susceptible, the distribution $f_I$ of the population of infected individuals depends on the full vector state, 
\rev{
\be\label{fI}
f_I= f_I(\au,t)= f_I(\ax,\av,t)= f(\ax,\av,t) , \quad\av\neq \ao ,
\ee
and globally we recover $f=f_S+f_I$.}

The knowledge of the functions $f_J(\au,t)$, $J \in \{S,I\}$ allows to compute all relevant observable quantities. In particular, the distributions at time $t \ge 0$ of the socio-physical states of the sub-populations of susceptible and infected people are given by the integrals
\be\label{socio}
F_J(\ax, t)=\int_{\mathbb{R}_+^2}f_J(\ax,\av,t)\,d\av,\quad J \in \{S,I\},
\ee
\rev{and considering the whole population we have $F(\ax, t)=F_S(\ax,t)+F_I(\ax,t)$.}
Hence, the integrals
\be\label{mass}
J(t)=\int_{\mathbb{R}_+^2}F_J(\ax,t)\,d\ax,\quad J \in \{S,I\}  
\ee
 denote the mass fractions at time $t \ge 0$ of the \rev{compartments $S$ and $I$, which respect the conservation of the total population, being $N=S(t)+I(t)$ constant in time.
Analogously, for $J(t)\neq0$ (i.e., only for non-empty compartments)} one can compute moments, defined, for any given constant  $\gamma>0$,  by 
\be\label{Xmom}
m_{J,i}^{\gamma}(t)= \frac 1{J(t)}\int_{\mathbb{R}_+^2}x_i^\gamma F_J(\ax,t)\,d\ax, \quad J \in \{S,I\},\, i =1,2.
\ee
The mean values, corresponding to $\gamma=1$,  are denoted by $m_{J,i}(t)$, $J \in \{ S,I\}$, $i =1,2$.
Clearly, choice $i=1$ corresponds to assessing, at time $t \ge 0$, the average resistance to the disease of the classes, while choice $i=2$ gives the average sociality of the classes.
 
On the pandemic side, the distribution of the virus in the \rev{classes reads
\be\label{statoV}
P_J(\av, t)=\int_{\mathbb{R}_+^2}f_J(\ax,\av,t)\,d\ax, \quad J \in \{S,I\},
\ee
and the distribution of the virus with respect to the whole population will be $P(\av, t)=P_S(\av, t)+P_I(\av, t)$. Notice here that, due to definition \eqref{fS}, $P_S(\av,t)=S(t)\cdot\delta(\av = \ao)$.}
As before, one can compute moments through the formula
\rev{
\be\label{Vmom}
n_{J,i}^\gamma (t) =  \frac 1{J(t)}\int_{\mathbb{R}_+^2}v_i^\gamma P_J(\av,t)\,d\av, \quad J \in \{S,I\},\, i =1,2.
\ee}
The mean values \rev{$n_{J,i}(t)$}, $J \in \{S,I\}$, $i =1,2$, corresponding to $\gamma=1$, give the average contagiousness of the virus ($i=1$) and its average severity ($i=2$) \rev{in each sub-population} at time $t \ge 0$.

The evolution of the densities $f_J$, $J \in \{S,I\}$, replicates the Boltzmann equation \eqref{Boltzmann}, with the additional subdivision of the population in compartments.
\rev{
From \eqref{Boltzmann}-\eqref{collision} we have indeed
\begin{equations}\label{Boltzmann-SI}
\frac{\partial f(\ax,\av,t)}{\partial t} &=\int_{\R_+^4} \kappa_*(\ax_*,\ay_*,\av_*,\aw_*)\Big(f_{S}(\ax_*,\av_*,t) + f_{I}(\ax_*,\av_*,t)\Big)\Big(f_{S}(\ay_*,\aw_*,t) + f_{I}(\ay_*,\aw_*,t) \Big) \,d\ay d\aw\\
&- \Big(f_{S}(\ax,\av,t) + f_{I}(\ax,\av,t)\Big)\int_{\R_+^4}\kappa(\ax,\ay,\av,\aw)  \Big( f_{S}(\ay,\aw,t) + f_{I}(\ay,\aw,t)\Big)\,d\ay d\aw\\
&= Q(f_S,f_S)+Q(f_S,f_I)+Q(f_I,f_S)+Q(f_I,f_I),
\end{equations}
where we have only used the definition $f=f_S+f_I$ in the r.h.s. of the identity.} 

\rev{It is natural to assume that the binary dynamic \eqref{eq:bin} modifying the social or the viral features satisfies
\[
Q(f_S,f_S)=0,\quad Q(f_I,f_S)=0,\quad Q(f_I,f_I)=0,
\] 
since it is only the interaction operator $Q(f_S,f_I)$ which propagates the pandemic due to the passage of susceptible individuals to the class of infected ones. It should be noted that, in principle, social characteristics could also have been influenced during these interactions that do not change the characteristics of the epidemic. For example, two susceptible individuals with different social characteristics interacting with each other may change the relative social attitudes. In order not to complicate the modelling unnecessarily, here we neglect these aspects that can be included following the ideas in^^>\cite{Esurvey, DPaTZ, DPTZ}. 
}

\rev{As a consequence, the contact function $\kappa$, governing the occurrence of the interactions between susceptible and infected, only depends on $(x_1,y_1,x_2,y_2,w_1,w_2)$ since the susceptible class is defined only when $v_1=0$ and $v_2=0$. Furthermore, we then assume that the contact function depends neither on the individuals' personal rate of resistance, $x_1$ and $y_1$, nor on the severity of the virus $w_2$.
Thus, in agreement with^^>\cite{Esurvey, DPaTZ, DPTZ}, we set the contact function $\kappa=\kappa(x_2,y_2,w_1)$ to be a non-negative function growing both with respect to the social activities $x_2$ and $y_2$ of the interacting individuals and to the degree of infectiousness $w_1$ of the virus present in the infected agent.
Moreover, we choose a contact function such that $\kappa(x_2,y_2,w_1) = 0$ when $x_2 = 0$, $y_2 = 0$ or $w_1 = 0$,
which expresses the fact that the epidemic cannot spread either in the absence of social contact between individuals or in the absence of contagiousness of the virus.}

A leading example of a contact function that fulfills the above requirements is obtained by choosing, for given positive constants $\delta,\eta,\theta$,
\be\label{kappa}
\kappa(x,y,w) = \theta x^\delta y^\delta w^\eta.
\ee
This choice corresponds to take the incidence rate proportional to the product of the number of contacts of $S$ and $I$ people and to the infectiousness of the virus present in the second agent. This generalizes a similar choice previously made in^^>\cite{Esurvey,DPaTZ,DPTZ}.

\rev{Hence, we finally obtain}
\be
\label{Boltzmann-SI1}
\frac{\partial \rev{f(\ax,\av,t)}}{\partial t} = K_I(f_S,f_I)(\ax,\av,t) - K_S(f_S,f_I)(\ax,\av,t),
\ee
in which we have renamed \rev{$Q^+(f_S,f_I)=K_I(f_S,f_I)$, $Q^-(f_S,f_I)=K_S(f_S,f_I)$, and}
\be\label{inci-I}
\rev{K_I(f_S,f_I)(\ax,\av,t) = \int_{\R_+^4} \kappa_*(x_{2,*},y_{2,*},w_{1,*}) f_{S}(\ax_*,\av_*,t) f_{I}(\ay_*,\aw_*,t)\,d\ay d\aw,}
\ee
\be\label{inci}
K_S(f_S,f_I)(\ax,\av,t) = f_{S}(\ax,\av,t)  \int_{\R_+^4}\rev{\kappa(x_{2},y_{2},w_{1})} f_{I}(\ay,\aw,t)\,d\ay d\aw.
\ee
\rev{Finally, by splitting the dynamics of susceptible individuals from those of the infected to obtain a system of equations in line with classical compartmental modelling approaches of 
epidemics, we get}
\begin{equations}
\label{Boltzmann-SIsyst}
\frac{\partial f_S(\ax,\av,t)}{\partial t} &= - K_S(f_S,f_I)(\ax,\av,t),\\
\frac{\partial f_I(\ax,\av,t)}{\partial t} &= K_I(f_S,f_I)(\ax,\av,t).
\end{equations}

In epidemiological modelling, the functions \rev{$K_S$ in \eqref{inci} and $K_I$ in \eqref{inci-I}} are usually known as the local incidence rates, which quantify the transmission of infection. Unlike the classical SIR model, where $K_S = K_I$, due to the presence of the state variable $\av$, these two functions take a different form from each other.
The function \rev{$K_S(f_S,f_I)(\ax,t)$} quantifies the amount of individuals that move from the compartment of susceptible people to the compartment of infected people, \rev{hence it represents the loss term of the interaction dynamics, while the amount of individuals entering into the infectious compartment is governed by $K_I(f_S,f_I)(\ax,t)$, being the gain term of the interaction}. 


In what follows, to maintain as much as possible the connection with the methods widely used in classical kinetic theory of rarefied gases, we will often resort to the weak formulation of the incidence functions $K_S(f_S,f_I)$ and $K_I(f_S,f_I)$, which corresponds to quantify their action on a given smooth function $\varphi(\ax,\av)$ (the observable function)^^>\cite{PTbook}. Starting from equation \eqref{inci}, this leads to the identity
\begin{equations}\label{inci-weak}
&\int_{\R_+^4}\va(\ax,\av) K_S(f_S,f_I)(\ax,\av,t)\, d\ax d\av  = \\ &\int_{\R_+^8}\rev{\kappa(x_2,y_2,w_1)}\va(\ax,\av) f_S(\ax,\av,t) f_I(\ay,\aw, t) \,d\ax d\ay d\av d\aw .
 \end{equations}
Instead, from equation \eqref{inci-I}, \rev{using the transformations of variables \eqref{k*} and \eqref{jacobian}, we have}
 \begin{equations}\label{inci-weakI}
&\int_{\R_+^4}\va(\ax,\av) \,K_I(f_S,f_I)(\ax,\av, t)\, d\ax  d\av  = \\
&\rev{\int_{\R_+^8} \kappa_*(x_{2,*},y_{2,*},w_{1,*})\va(\ax,\av)\,f_S(\ax_*,\av_*,t)f_I(\ay_*,\aw_*, t)  \,d\ax d\ay d\av d\aw =} \\
&\rev{\int_{\R_+^8} \kappa(x_{2,*},y_{2,*},w_{1,*})\va(\ax,\av)\,f_S(\ax_*,\av_*,t)f_I(\ay_*,\aw_*, t)  \,d\ax_* d\ay_* d\av_* d\aw_* =} \\
&\int_{\R_+^8} \rev{\kappa(x_{2},y_{2},w_{1})}\va(\ax^*,\av^*)\,f_S(\ax,\av,t)f_I(\ay,\aw, t)  \,d\ax d\ay d\av d\aw.
 \end{equations}
\rev{Note that the last equality is simply obtained by renaming the integration variables. 
Consequently, the state vector $\au^*=(\ax^*,\av^*)$ is the post-interaction state of an individual that got infected interacting with an infectious one, resulting from the pre-interaction pair $\au=(\ax,\av)$ and $\aut=(\ay,\aw)$ (in this specific case $\av=\ao$, being the first agent a susceptible), exactly as the state vector $\au=(\ax,\av)$ is the post-interaction state resulting from the pre-interaction pair $\au_*=(\ax_*,\av_*)$ and $\aut_*=(\ay_*,\aw_*)$. Please refer again to Figure \ref{fig:transform} for a schematic representation of these transformations.
} 

Within this picture, we will assume that in any new infection the presence of the virus modifies the socio-physical conditions of the susceptible individual and, at the same time, the viral impact of the new infected individual can be different from the viral impact of the infectious individual responsible for contagion.

Finally, it is important to underline that, in presence of a constant observable function \rev{$\va(\ax,\av)=\va(\ax^*,\av^*)=C$, from \eqref{inci-weak} and \eqref{inci-weakI}} we have the equality
\be\label{a1}
\int_{\R_+^4}K_S(f_S,f_I)(\ax,\av, t)\, d\ax d\av = \int_{\R_+^4}K_I(f_S,f_I)(\ax,\av, t)\, d\ax d\av,
 \ee
which confirms that the total number \rev{of individuals $N$} of the population is preserved.



\subsection{\rev{Extension to SIR modelling}}
As seen so far, the kinetic dynamics presented involves only subjects belonging to the $S$ and $I$ compartments. To also take into account the mechanism of healing (or death) of infected individuals, we extend the model just introduced according to the SIR dynamics^^>\cite{KM}.
\rev{To this aim, we introduce an additional compartment of ``removed'' individuals (R), who either recovered from the disease and have become immune to it or have deceased.}
To maintain memory of the fact that recovered individuals were infected with the virus, their distribution is denoted by 
\rev{\be
f_R =f_R(\au, t)=f_R(\ax, \av, t) = f(\ax,\av,t)\cdot \delta(v_1 = 0),\quad v_2\neq 0.
\ee}
In other words, recovered individuals are a subset of the previous compartment of infected characterized by an infectiousness equal to zero.
\rev{As a consequence, in this modelling framework the definition of the distribution $f_I$ changes into
\be
f_I= f_I(\ax,\av,t)= f(\ax,\av,t) , \quad v_1 \neq 0,
\ee
so that we recover $f=f_S+f_I+f_R$. Clearly, definitions \eqref{socio}, \eqref{mass}, \eqref{Xmom}, \eqref{statoV} and \eqref{Vmom} directly apply also for the compartment $R$.}

It is important to remark that the value $v_2$ could be fruitfully used to measure the percentage of infectious people who do not survive to the disease. This can be obtained by imposing an upper bound $\hat v_2$ above which the infectious individual do not move to the compartment of recovered people, but do move into the compartment of the deceased (D). Therefore, \rev{in an SIRD compartmentalization} recovered individuals will be characterized by a value $\rev{\av} =(0, v_2)$, in which $0<v_2 < \hat v_2$.

\rev{Note that the healing process is not related to the binary interactions between agents, i.e. it does not change the modelling conclusions of the previous section, so it is necessary to include the recovery dynamics as a new additional effect in the model. 
To this aim, we include a recovery process in the Boltzmann-type contagious model \eqref{Boltzmann-SIsyst} in analogy with the classical SIR model in literature^^>\cite{KM}. Thus, the SIR kinetic model here proposed results defined by the system of integro-differential equations}
\begin{equations}\label{sir-gamma}
\frac{\partial f_S(\ax,\av,t)}{\partial t} &= -K_S(f_S,f_I)(\ax,\av,t) ,
\\
\frac{\partial f_I(\ax,\av,t)}{\partial t} &= K_I(f_S,f_I)(\ax,\av, t)  - \gamma(\ax,\av) f_I(\ax,\av,t),
\\
\frac{\partial f_R(\ax,\av, t)}{\partial t} &= \gamma(\ax,\av) f_I(\ax,\av,t).
\end{equations}
Here, the function $\gamma(\ax,\av) = \gamma(x_1,v_2) >0$ represents the recovery rate, which is assumed to be dependent on the joint action of the personal resistance $x_1$ of the infected individual and the severity $v_2$ of the disease.
Let us notice that, if both the contact function $\kappa$ and the recovery rate $\gamma$ are assumed to be constant, integration of system \fer{sir-gamma} shows that the mass fractions, as given by \fer{mass}, satisfy the classical SIR evolution^^>\cite{KM}.

\rev{
\begin{remark}
More in general, it would be interesting to derive a macroscopic dynamics in the case of non constant contact functions by integrating with respect to the vectors $\ax$ and/or $\av$, to obtain evolutionary equations for the moments. For example, taking moments with respect to social characteristics would result in a model for viral dynamics that can be related to previous models in the literature, such as the one presented in^^>\cite{DLT23}.
However, this would require the introduction of a closure to be made with respect to the socio-physical characteristics of individuals. This can be done by computing an approximate equilibrium state of the social characteristic following the mean-field analysis in \cite{PTbook,Esurvey,PT}, or simply by assuming a mono-kinetic closure as in^^>\cite{DLT}. Although this may lead to simplified mathematical models which can be more easily studied and integrated numerically, here we would like to keep the main focus on the binary interaction process and its specific construction through an utility process. Therefore we limit ourselves to present the temporal evolution of the moments in the numerical tests discussed in Section \ref{numtest} and leave possible derivation of the macroscopic dynamics to future research.
\end{remark}
}

\section{Virus-agent interaction based on mutual utility}
In general, \rev{both the SI-type model \eqref{Boltzmann-SIsyst} and} the SIR-type model \fer{sir-gamma} are characterized by an \emph{inhomogeneous mixing}. The complete description of these models then requires the knowledge of the binary interaction $(\ax,\aw) \to (\ax^*,\rev{\aw^*})$, \rev{i.e., of the functions $\Phi$ and $\Psi$ in \eqref{eq:bin}}, which characterizes the contagion of a susceptible individual with socio-physical condition $\ax$  by an infectious person with viral impact $\aw$, making him/her infected with modified socio-physical condition $\ax^*$ and viral impact $\rev{\aw^*}$. 
This is an extremely difficult problem which includes, from the biological point of view, the understanding of the reasons why the dynamics at the molecular scale leads to an evolution in time of the biological features of the virus up to mutations, and how these affect the in-host consequences of the presence of the virus.
Having this behaviour in mind, \rev{to define the post-interaction states of the dynamics under study,} we will hypothesize, in a largely arbitrary manner, that both individual's and virus' behaviour are driven by utility reasons, mainly related to their survival. Once this hypothesis has been fixed, a well-known interaction rule which is based on utility functions can be identified in the Edgeworth box, of common use in economics. As this topic may be unfamiliar to those who are involved in compartmental epidemiology, we will briefly present it in Section \ref{Edge}, \rev{before discussing the resolution of the problem of post-interaction states in detail in Section \ref{Edge_SIR}}.


\subsection{Edgeworth box and Cobb-Douglas utility function}\label{Edge}
People exchange goods. The advantages they obtain are contingent upon the extent and conditions of their exchange. This important question is attempted to be answered by price theory. A binary transaction might involve a wide range of trades, some of which would be preferred by one party over the other but all of which would be advantageous to both.  In the following, we will assume that the agents inside the system own the same two kinds of goods and that there is no production, meaning that the overall quantity of goods stays constant.
If $(x_A, y_A)$ and $(x_B, y_B)$ denote the quantity of goods of two agents $A$ and $B$, respectively, then
 \be\label{per}
p_A = \frac{x_A}{x_A + x_B}, \quad q_A = \frac{y_A}{y_A + y_B}
 \ee
are the percentages of goods of the first agent. As a matter of definitions, the point $(p_A,q_A)$ belongs to the square $\mathcal{S}= [0,1]\times[0,1]$. 

To provide a foundation for the motivations behind trading, conventional wisdom holds that an agent's actions are determined by its utility function. The \emph{Cobb-Douglas utility function} is among the most often used of these functions:
 \be\label{CD}
 U_{\a,\b}(p,q) = p^\alpha q^\beta, \quad \alpha + \beta = 1.
 \ee
Each agent aims to optimize its individual satisfaction through trade. The parameters $\alpha$ and $\beta$ are associated with the preferences an agent attributes to the two goods. When $\alpha > \beta$, the agent has a preference for acquiring goods of the first type (designated by $x$). If $\alpha = \beta = 0.5$, it is evident that both goods hold equal importance for agent $A$. Given the percentage point $(p_A,q_A)$ of agent $A$, the curve
 \[
U_{\a,\b}(p,q) = U_{\a,\b}(p_A,q_A)
 \]
denotes the \emph{indifference curve} for agent $A$. In fact, any point on the indifference curve yields the same level of utility for agent $A$. It is important to observe that the indifference curve for $A$ lies entirely within set $\bigS$ and divides the square into two distinct regions:
 \[
U^-_A = \left\{(p,q) :U_{\a,\b}(p,q) < U_{\a,\b}(p_A,q_A)\right\},
\]
\[
 U^+_A =
\left\{(p,q) :U_{\a,\b}(p,q) > U_{\a,\b}(p_A,q_A)\right\}.
 \]
Evidently, any trade that shifts the proportions of agent $A$ into the region $U^+_A$ will enhance the utility function and be deemed acceptable by the agent.

In the scenario of a binary trade, agent $B$ also possesses a Cobb-Douglas utility function, with preference parameters generally distinct from those of agent $A$. Similarly, there exists an indifference curve for agent $B$ and a specific region where the utility of $B$ increases post-trade. Ultimately, a trade becomes acceptable for both agents when their respective percentages of goods after the trade fall within the regions where both utility functions experience an increase.

\begin{figure}[t]
\centering
\includegraphics[width=0.48\textwidth]{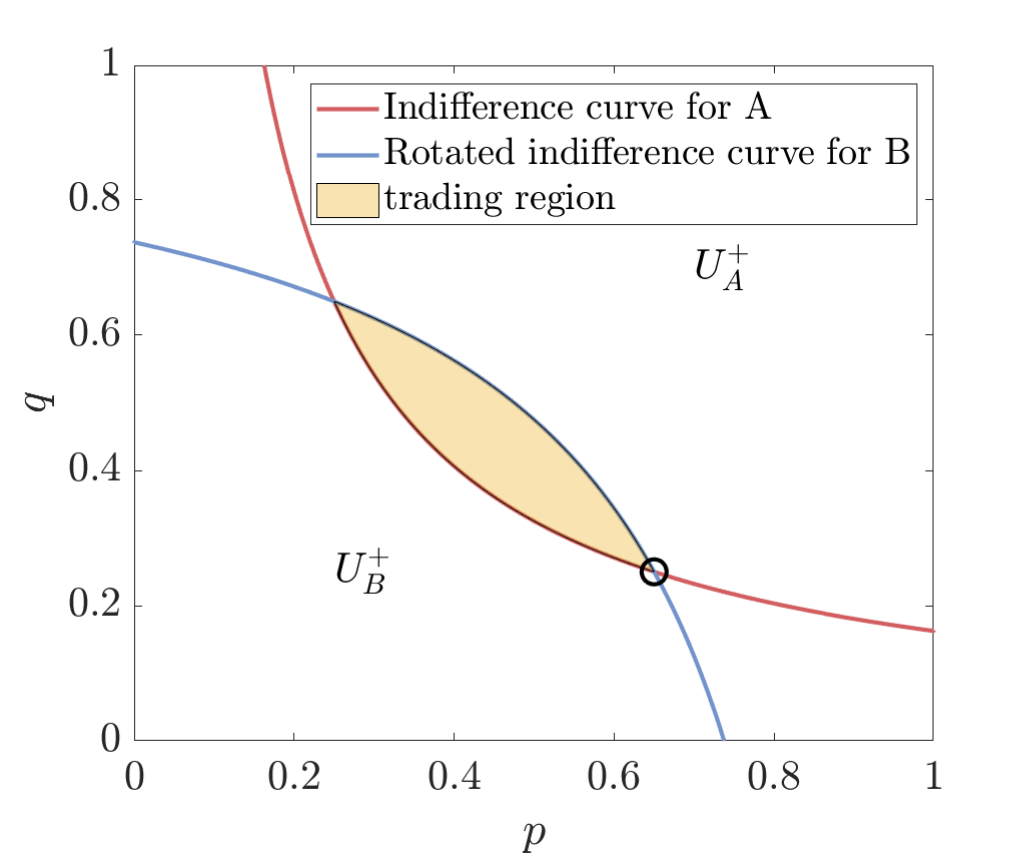}
\includegraphics[width=0.48\textwidth]{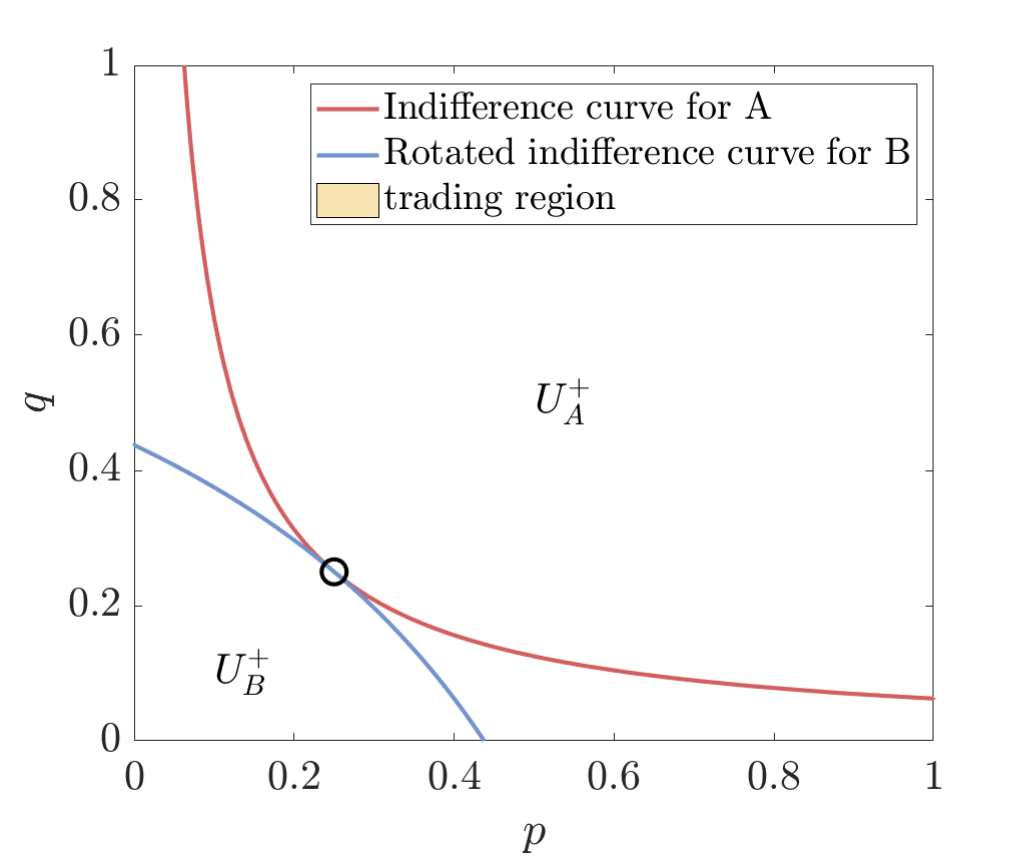}
\caption{Edgeworth box. Indifference curves in a binary trade for the Cobb-Douglas utility function \eqref{CD} with $\alpha=\beta=0.5$. On the left, the trading region for $(p_A,q_A)=(0.65,0.25)$ (and, therefore, $(p_B,q_B)=(0.35,0.75)$), indicated with the black circle; on the right, there is a unique tangent point (Pareto optimal) at $(p_A,q_A)=(0.25,0.25)$.}
\label{fig:CD}
\end{figure}

An insightful perspective on such a scenario is offered by the Edgeworth Box, named after Francis Y. Edgeworth, the author of the 19th-century work \emph{Mathematical Psychics}^^>\cite{Edg}. The Edgeworth box is constructed by rotating the square $\bigS$ where the indifference curve of agent $B$ is delineated by $180^\circ$ around the centre of the square $(0.5,0.5)$ and by considering the indifference curve of agent $A$ and the rotated indifference curve of $B$ together on the same square (see Figure \ref{fig:CD} for an example).

Any point within the Edgeworth box represents a potential distribution of the total quantities of $A$ and $B$, where the share of $A$ is measured from the lower left-hand corner, and the share of $B$ is measured from the upper right-hand corner. Each conceivable trade corresponds to a transition from one point in the box to another.
By design, given the jointly concave nature of the Cobb-Douglas utility function, the utility of $A$ increases when the point shifts up and to the right, while the utility of $B$ increases when the point moves down and to the left. The specific region where this occurs is precisely the overlapping area between $U^+_A$ and $U^+_B$, rotated by $180^\circ$.

The operational approach to trading facilitated by the Edgeworth box yields intriguing implications. When the percentages of $A$ and $B$ allow for a potential trading region, the completion of a trade causes the point to shift within the region, subsequently reducing the scope of the new trading area. If both agents intelligently select a point within the region where the two indifference curves are tangent, indicating a point of maximum mutual satisfaction, any movement from this point towards a higher indifference curve for one agent corresponds to a lower curve for the other. Consequently, any trade that benefits one agent harms the other. The points where the two curves are tangent are not unique, and the collection of points from which no further mutually beneficial trading is feasible, known as \emph{Pareto optimal} points, constitutes the \emph{contract curve}.
The contract curve has a connection to prices. At the point of tangency between the two indifference curves, the slope of the tangency line denotes the relative prices of the two goods. Therefore, there exist relative prices consistent with Pareto optimality, and these prices maximize the potential budget for both agents.

A kinetic model of Boltzmann type, with binary interactions based on Edgeworth box, has been presented and discussed in^^>\cite{TBD}. In this model, a key aspect is the presence of randomness in the trade, motivated by the incomplete knowledge of the market. 

\subsection{A viral interaction model}\label{Edge_SIR}
\emph{Mutatis mutandis}, the modelling assumptions in^^>\cite{TBD} can be fruitfully applied to construct a \emph{viral} interaction between the pair of susceptible and infected individuals, based on mutual utility. 

Let $S$, respectively $I$, denote a susceptible, respectively infectious, individual, which are identified by the pair $(\ax,\aw)$ of the socio-physical conditions $\ax$ of $S$ and the viral impact $\aw$ of $I$. In what follows,
we assume that the individual  $S$ and  the virus present in $I$  exhibit the same type of utility
function \fer{CD}, but in general with different preferences. We assume moreover that the preferences of $S$ are determined by the search of its own best socio-physical conditions, while the preferences  of the virus present in $I$ are linked to its survival. 
We outline that possible heterogeneity in individuals' preferences is neglected in the model, an acceptable simplification because all individuals are exposed to similar risk. Likewise, we neglect possible heterogeneity in virus preferences. Clearly, in both classes, preferences can be assumed to vary with time without any additional difficulty in modelling. 

Before entering into the detail of the interaction, let us fix some points that allow to better identify the various parameters characterizing  the \emph{virus--agent} Edgeworth box.  It is reasonable to assume that the evolutionary choice of the virus goes toward increasing contagiousness ($w_1$), since high contagiousness ensures its survival. Conversely, individuals will aim at maintaining a higher level of survival rate ($x_1$). Therefore, the higher the contagiousness of the virus, the more the disease resistance (survival rate) of the susceptible individual will decrease, and vice versa. Additionally, we can assume that the individual behaviour, and the consequent evolutionary preferences, are heavily dependent on the state of the epidemic spread. This can be expressed by hypothesizing that in presence of a high degree of virus severity ($w_2$), individuals' sociality tends to be reduced by them ($x_2$), while the opposite effect occurs in light of a low degree of severity, since in this case the risks associated with social activity are perceived as low.

According to this discussion,  it seems appropriate to relate the pairs  \emph{resistance--contagiousness} and \emph{sociality--severity},  which are associated to something similar to a conservation law, namely that the quantities \emph{resistance$+$contagiousness} and \emph{sociality$+$severity} are invariants of the interaction, and, furthermore, that they are related to similar degrees of preferences. In this sense, during the evolutionary dynamics, when one variable decreases, its pair counterpart increases, and vice-versa. Thus, the Edgeworth box pair $(p_S,q_S)$ of the susceptible individual  is  defined by
 \be\label{perS}
p_S = \frac{ x_1}{ x_1 +  w_1}, \quad q_S = \frac{ x_2}{ x_2 +  w_2},
 \ee
while the pair $(p_I,q_I)$ of the virus in any infected individual is 
 \be\label{perI}
p_I = \frac{ w_1}{ x_1 + w_1}, \quad q_I = \frac{ w_2}{ x_2 +  w_2}.
 \ee
To proceed, it is necessary to explicitly identify a possible virus-agent interaction that leads to an increase in the utility functions of both. The underlying idea is that the time evolution of the solution of the system \fer{sir-gamma} is essentially characterized by the fact that the elementary interactions  lead to the growth of the Cobb-Douglas utility functions. To this aim, we extend the analysis of^^>\cite{TBD} to the case where the  individuals are endowed with different preferences. 

Following the analysis  in^^>\cite{TBD},   if an  individual, denoted by $A$,  has percentages $p$ and $q$ and preferences $(\alpha,\beta)$, with $\alpha +\beta= 1$, in its
Edgeworth box, we assume that the interactions are characterized by the movements of the
point $(p,q)$ into $(p^*,q^*)\in \bigS$, where
\begin{equations}\label{tr}
&p^* = p + \mu\beta(q-p) \\
    &q^* = q +  \tilde\mu\alpha(p-q).
\end{equations}
We remark that in this case the variation of the position $(p,q)$ of the individual is fully characterized by its own preferences. In \fer{tr}   $\mu$ and $\tilde \mu$ are non-negative random
variables with  mean $\lambda>0$  and {finite variance}. We will assume moreover
that
 \be\label{co3}
 \mu\beta < 1, \quad  \tilde\mu\alpha < 1.
 \ee
Under this condition, the post-interaction point $(p^*, q^*)$ belongs to
$\bigS$, and it is an admissible point for the Edgeworth box. It is
immediate to show that, unless $p=q$, and in absence of randomness, that is
if the pair $(p^*,q^*)$ is given by 
\begin{equations}\label{tr3}
&p^* = p + \lambda\beta(q-p) \\
    &q^* = q +  \lambda\alpha(p-q),
\end{equations}
 the interaction, for $\lambda <1$, {increases the Cobb-Douglas utility function} of the individual.
Indeed, for $\lambda <1$, a simple computation gives^^>\cite{TBD}
\begin{equation}\label{qua}
\frac d{d\lambda} U_{\a,\b}( p^*, q^* ) =\alpha\beta\left(1-\lambda\right)(p-q)^2 (p^*)^{\alpha-1}(q^*)^{\beta-1} > 0.
\end{equation}
Hence, if both individuals share the same preferences, interactions of type \fer{tr} increase the utilities of both.

Let us now suppose that a second individual, denoted by $B$, has preferences $(\alpha_1,\beta_1)$, with $\alpha_1 +\beta_1=1$, and $\alpha_1\not=\alpha$, hence $\beta_1\not=\beta$.  Suppose moreover that
\be\label{pref1}
\frac\alpha\beta < \frac{\alpha_1}{\beta_1}.
\ee
At difference with the previous case, we  consider as possible interactions of the  individual $A$ also the movements of the
point $(p,q)$ into $(p_1^*,q_1^*)\in \bigS$, where
\begin{equations}\label{tr1}
&p_1^* = p + \mu\beta_1(q-p) \\
    &q_1^* = q +  \tilde\mu\alpha_1(p-q).
\end{equations}
In this case, the variation of the position $(p,q)$ of the individual $A$ is influenced by the preferences of the individual $B$.
For the sake of simplicity, we assume that the randomness in \fer{tr1}  is determined by the same random variables $\mu$ and $\tilde \mu$ appearing in  \fer{tr}, further  satisfying 
 \be\label{co31}
 \mu\beta_1 < 1, \quad  \tilde\mu\alpha_1 < 1,
 \ee
so that  the post-interaction point $(p_1^*, q_1^*)$ belongs to
$\bigS$. Different choices of the random effects could be considered, at the price of an increasing rate of computations. 

In absence of randomness, 
 the pair $(p_1^*,q_1^*)$ associated to \fer{tr1} is given by 
\begin{equations}\label{tr2}
&p_1^* = p + \lambda\beta_1(q-p) \\
    &q_1^* = q +  \lambda\alpha_1(p-q).
\end{equations}
In correspondence to  interaction \fer{tr2}, one obtains
\begin{equations}\label{der2}
\frac d{d\lambda} U_{\a,\b}( p_1^*, q_1^* ) = &(p_1^*)^{\alpha-1}(q_1^*)^{\beta-1}\left\{ (\beta- \beta_1\lambda) \alpha_1 p^2  + \right.\\ 
& \left. (\alpha- \alpha_1\lambda) \beta_1 q^2  +2\left[\alpha_1\beta_1 \lambda - \frac 12(\alpha\beta_1 +\beta\alpha_1) \right]pq \right\}.
\end{equations}
Let $\lambda$ satisfy the condition
\be\label{newb}
 \lambda \le  \frac\alpha{\alpha_1},
 \ee
so that, in view of \fer{pref1}, $\lambda < \beta/{\beta_1}$,
and let us set $x = p/q$. 
Under  condition \fer{newb}, the second-degree polynomial
 \[
 (\beta- \beta_1\lambda) \alpha_1 x^2  + 
   2\left[\alpha_1\beta_1 \lambda - \frac 12(\alpha\beta_1 +\beta\alpha_1) \right]x +  (\alpha- \alpha_1\lambda) \beta_1 
 \]
is non-negative if and only if $x$ lies outside the interval of the two roots $x_1$ and $x_2$, which are given by
 \[
 \quad x_1 = \frac{(\alpha- \alpha_1\lambda) \beta_1}{ (\beta- \beta_1\lambda) \alpha_1}, \quad x_2=1.
 \]
Since \fer{pref1} holds, $0 < x_1 < x_2$. Hence, for all values of $\lambda$ satisfying \fer{newb}, the utility function $U_{\a,\b}( p_1^*, q_1^* ) $ is non-decreasing in the half-square $p\ge q$. 

Thus, if \fer{pref1} holds, and $p \ge q$, for any given $\lambda$ satisfying  \fer{newb},
\[
U_{\a,\b}( p_1^*, q_1^* ) \ge U_{\a,\b}( p, q ), \quad U_{\a_1,\b_1}(1- p_1^*,1- q_1^* ) \ge U_{\a_1,\b_1}( 1-p,1- q ),
\]
and the utility functions of both individuals are non-decreasing.

The previous computations can be repeated in the half-square $p \le q$ with the interaction \fer{tr3}, and  utility function $U_{\a_1,\b_1}( p^*,q^* )$. In this case, provided $\lambda$ satisfies the bound
\be\label{newb1}
 \lambda \le  \frac {\beta_1}\beta,
\ee
so that, in view of \fer{pref1}, $\lambda<\alpha_1/{\alpha}$,  the utility functions of both individuals are non decreasing. 

Hence, joining the two results, we conclude that, in absence of randomness, for any given pairs of preferences such that \fer{pref1} holds, and for any $\lambda$ such that 
\be\label{newb2}
 \lambda \le \min\left\{\frac\alpha{\alpha_1}; \frac {\beta_1}\beta\right\},
\ee
the interaction defined by \fer{tr3} in the half-square $p\le q$ (respectively \fer{tr2} in the half-square $p \ge q$ ), is such that the Cobb-Douglas utility functions of both individuals are non-decreasing. 

 The just described interaction refers to the purely theoretical situation in which the individual  knows all about the interaction and its result. In general, this is not realistic, and it appears 
reasonable to postulate  that the result of the interaction  has a share of unpredictability, so that increasing of utility can be assumed only in the mean. If one agrees with this, then interactions of type \fer{tr} satisfy all constraints of Edgeworth box, but the single interaction can move the point in a region which is convenient only for one of the two interacting agents, as well as in the region which is not convenient for both. 

Let $I_E(x)$ denote the characteristic function of the set $E$, namely the function such that
\[
I_E(x) = 1 \quad {\rm if} \,\,\, x \in E; \qquad I_E(x) = 0 \quad {\rm if} \,\, \, x \notin E.
\]
Then, if we define
\be\label{short}
 \Psi_\lambda (a,b,p,q) = \lambda \left[ a I_{\{p<q\}} + b I_{\{p>q\}}\right] ,
\ee
we assume that the Cobb-Douglas increasing interaction characterizing the pairs of preferences $(\alpha_S, \beta_S)$ and $(\alpha_I, \beta_I)$ is given by
\begin{equations}\label{short1}
&p^* = p +   \Psi_\mu (\beta_S,\beta_I,p,q) (q-p) \\
    &q^* = q + \Psi_{\tilde\mu} (\alpha_S,\alpha_I ,p,q)(p-q),
\end{equations}
The interaction \fer{short1} is further characterized by the random variables $\mu$ and $ \tilde\mu$ of mean $\lambda$ satisfying \fer{newb2}, and such that both \fer{co3} and \fer{co31} hold.

Going back to the object of our analysis,  the interaction modifies the pairs  $(p_S,q_S)$, $(p_I, q_I)$ of the susceptible individual and, respectively, of the virus according to 
\begin{equations}\label{contact}
&p_S^* = p_S+  \Psi_\mu (\beta_S,\beta_I,p_S,q_S) (q_S-p_S) \\
    &q_S^* = q_S +    \Psi_{\tilde\mu} (\alpha_S,\alpha_I ,p_S,q_S)(p_S-q_S)\\
   &p_I^* = p_I+   \Psi_\mu (\beta_I,\beta_S,p_I,q_I) (q_I-p_I) \\
    &q_I^* = q_I +  \Psi_{\tilde\mu} (\alpha_I,\alpha_S ,p_I,q_I) (p_I-q_I).
\end{equations}

Note that the post-interaction values in \fer{contact}  satisfy the obvious relations
\[
p_S^*+p_I^* = q_S^*+q_I^* =1,
\] 
which guarantee that the post-interaction values are percentage values split between $S$ and $I$, like in \fer{perS} and \fer{perI}.

Moreover, note that the knowledge of the percentages $p_S^*$ and $p_I^*$ (respectively $q_S^*$ and $q_I^*$) does not allow to recover uniquely the post-interaction values $(\ax^*,\aw^*)$, since the relations \fer{perS} and \fer{perI} admit infinite solutions. 
However, assuming the  constraints $x_1 + w_1= x^*_1 + w^*_1$ and $x_2 + w_2=
x^*_2 + w^*_2$, we obtain that the interaction between the individual in $S$ with state $\ax$ and preferences $(\alpha_S,\beta_S)$ and the  virus in $I$  with state $\aw$ and preferences $(\alpha_I,\beta_I)$ is given by
\begin{equations}\label{trA}
  &x^*_1 = x_1 +  \Psi_\mu \left(\beta_S,\beta_I,\frac{x_1}{x_1+w_1}, \frac{x_2}{x_2+w_2}\right) \left( \frac{x_1+w_1}{x_2 + w_2} x_2 - x_1\right) \\
   & x^*_2 = x_2 +  \Psi_{\tilde\mu} \left(\alpha_S,\alpha_I,\frac{x_1}{x_1+w_1}, \frac{x_2}{x_2+w_2}\right)  \left( \frac{x_2 + w_2}{x_1+w_1} x_1 - x_2\right)
   \\
  &w^*_1 = w_1 +  \Psi_\mu \left(\beta_I,\beta_S,\frac{w_1}{x_1+w_1}, \frac{w_2}{x_2+w_2}\right)  \left( \frac{x_1+w_1}{x_2 + w_2} w_2 - w_1\right) \\
   & w^*_2 = w_2 + \Psi_{\tilde\mu} \left(\alpha_I,\alpha_S,\frac{w_1}{x_1+w_1}, \frac{w_2}{x_2+w_2}\right) \left( \frac{x_2 + w_2}{x_1+w_1} w_1 - w_2\right).
 \end{equations}

\begin{figure}[t]
\centering
\includegraphics[width=0.98\textwidth]{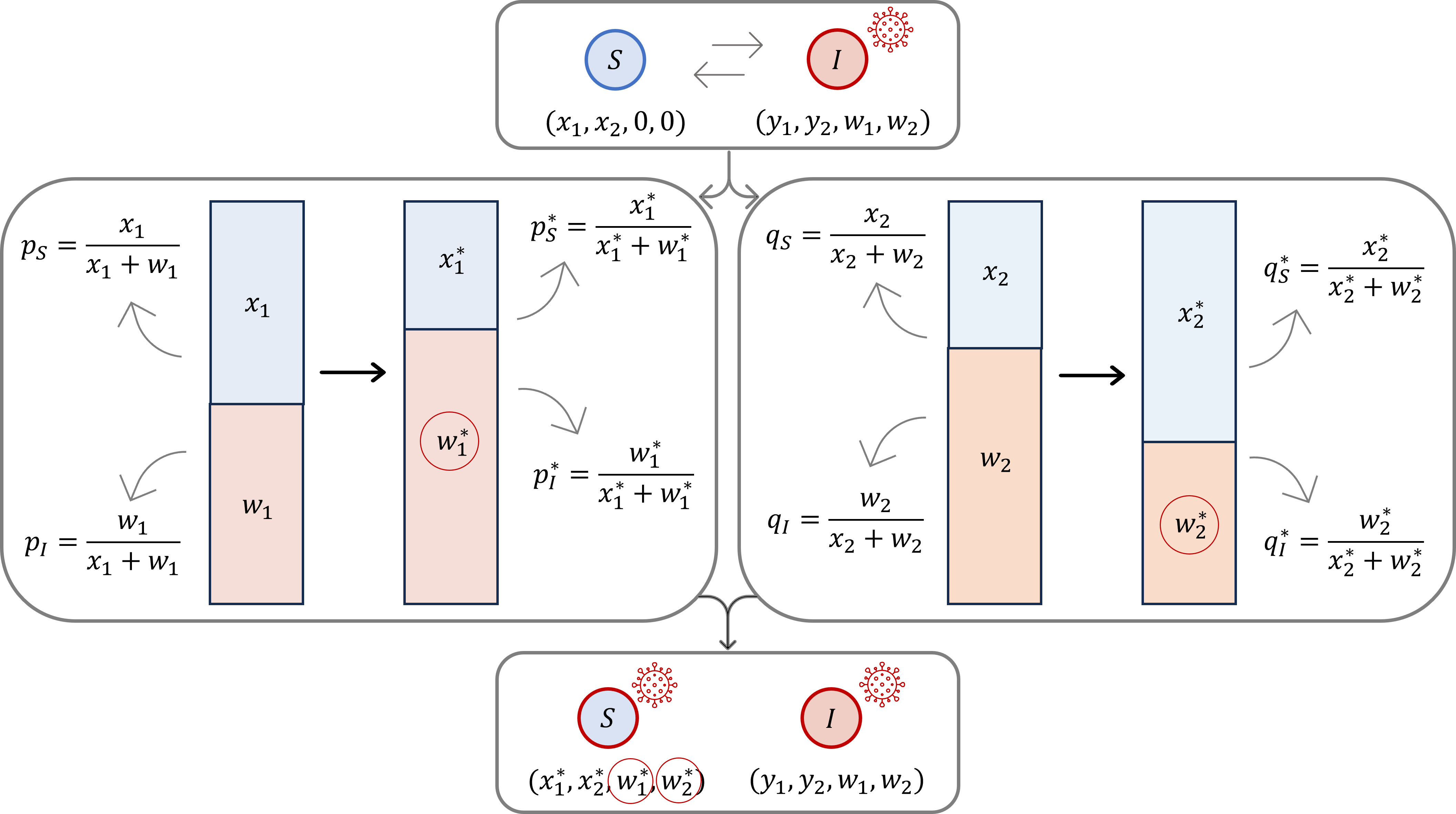}
\caption{Sketch of the interaction between the pair of susceptible ($S$) and infected ($I$) individuals (\emph{virus-agent} interaction) based on mutual utility with respect to the invariants \emph{resistance}+\emph{contagiousness} ($x_1+w_1$) and \emph{sociality}+\emph{severity} ($x_2+w_2$). The post-interaction socio-physical and viral states, $(x_1^*,x_2^*)$ and $(w_1^*,w_2^*)$, respectively, are both acquired at the level of the susceptible state update, whereas the pre-infected individual $I$ does not undergo a change of state.}
\label{fig:virus-agent}
\end{figure}
 
The process just described clearly follows an evolutionary dynamics between the virus (in the infected individual) and the susceptible agent that traces the Cobb--Douglas mechanism of continuous mutual utility. In fact, the post-interaction quantities in \fer{trA} are related to the pre-interaction quantities by non-linear relationships such that, at least on average, the Cobb--Douglas utility functions of both individuals increase: both the susceptible individual and the virus evolve by pursuing their own (distinct) interests.
From a purely epidemiological point of view, it is spontaneous to assume that the advantages gained by the virus in each evolutionary step do not manifest themselves in the updating of the $\aw$ utility variables proper to the $I$ individual, but rather in an activation of the viral variables of the $S$ subject. The latter, in fact, as a result of the interaction with $I$ (with virus that has followed an evolution that wishes to increase its contagiousness and severity) becomes the recipient of the increased utility share gained from the evolutionary process, expressed in terms of viral impact.
Thus, considering an interaction between $f_S(\ax,\av,t)$, with $\av = \ao$, and $f_I(\ay,\aw,t)$, we compute the update $\ax \to \ax^*$ and \rev{activate the null viral impact of the $S$ agent, through} $\av \to \av^* = \aw^*$, as in \fer{trA}.
At the same time, both socio-physical condition $\ay$ and viral impact $\aw$ of the agent $I$ remain unchanged at the end of the interaction. 
A sketch of this process is shown in Figure \ref{fig:virus-agent}.
This modelling choice finds its rationale also when observing that an infected individual does not tend to acquire greater infectiousness or severity on its own after an interaction with a susceptible; rather, the increase in viral disease characteristics occurs at the level of the interacting pair, with the susceptible becoming infected by activation of its viral load.

\begin{remark}
Interaction \fer{tr} is only one of the possible interactions that lead to an increased utility, but many others can be considered. However, interactions of type \fer{tr} have the interesting property to be linear in the percentage values entering into the Edgeworth box, and to furnish explicit values for the interactions which lead to the correct convenient area for both agents. This will be enough to perform a numerical study of the evolution in time of the virus-agent dynamics.
\end{remark}

\section{Numerical tests}\label{numtest}
To show the validity of the proposed model, we perform several numerical experiments employing one of the most commonly used Monte Carlo methods for the simulation of the interaction operators in \eqref{Boltzmann-SIsyst}, namely Nanbu-Babovsky's scheme. For the detailed algorithm the reader can refer to^^>\cite{PR,PTbook}. To numerically solve the complete SIR system \eqref{sir-gamma}, we apply a classical splitting procedure, for which the interaction operator and the recovery process are solved in sequence^^>\cite{PTbook}. 

\rev{
To present results obtained, we will make use of the following definitions of the marginal distributions at time $t \ge 0$ of the sole resistance to the disease ($F_1$) and of the sole sociality ($F_2$) of the population,
\be\label{socio_marg}
F_1(x_1, t)=\int_{\mathbb{R}_+}F(\ax,t)\,dx_2, \qquad F_2(x_2, t)=\int_{\mathbb{R}_+}F(\ax,t)\,dx_1,
\ee
and of the marginal distributions of the sole contagiousness ($P_1$) and sole severity ($P_2$) of the virus with respect to the totality of individuals,
\be\label{statoV_marg}
P_1(v_1, t)=\int_{\mathbb{R}_+}P(\av,t)\,dv_2, \qquad P_2(v_2, t)=\int_{\mathbb{R}_+}P(\av,t)\,dv_1.
\ee
Moreover, taking into account the entire set of agents to compute the mean resistance ($m_{1}$) and mean sociality ($m_{2}$) of the population, we have
\be\label{XmomTot}
m_{i}(t)= \frac 1{N(t)}\int_{\mathbb{R}_+^2}x_i^\gamma F(\ax,t)\,d\ax,\quad i =1,2,
\ee
while mean contagiousness ($n_1$) and mean severity ($n_2$) of the virus in the total population will be
\be\label{VmomTot}
n_{i}(t)= \frac 1{N(t)}\int_{\mathbb{R}_+^2}v_i^\gamma P(\av,t)\,d\av,\quad i =1,2,
\ee
}

\paragraph{Test 1.}
In the first numerical test, we consider a population of $N=10^5$ individuals in which, at the beginning of the simulation, presents only 1 infected agent, while the rest are susceptible. We define the same preference coefficients for all $S$ and $I$ agents, fixing $\alpha_S = \alpha_I = \beta_S = \beta_I = 0.5$, and characterize the uniformly distributed random variables $\mu$ and $\tilde\mu$ with the same mean $\lambda = 0.99$, satisfying \fer{newb2}, and same variance equal to $0.01$.
We adopt a simple setting in which the interaction kernel is constant, choosing first $\kappa = 1$ (case a) and then $\kappa = 0.5$ (case b), and we perform both test cases in two different configurations: considering the simplest possible compartmentalization of only S and I agents \rev{in \eqref{Boltzmann-SIsyst}} (thus, omitting the healing process) and, later, considering the classical SIR-type model \rev{in \eqref{sir-gamma}}, with $\gamma(\ax,\av) = 10^{-1}\, v_2/x_1$ days$^{-1}$.
For all the simulations we fix the same initial distributions for the state variables $\ax$ and $\av$. We consider a gamma distribution with shape and scale parameter respectively equal to 3 and 1, rescaled by a factor of 0.1, for both the socio-physical conditions $\ax=(x_1,x_2)$, and we attribute a random value of viral impact states $\av=(v_1,v_2)$ to the single infected agent following a Gaussian distribution with mean 0.4 and variance 0.01.
We run the simulation up to the final time $T = 60$ days in the case of the SI configuration and $T = 90$ days in the SIR one. We fix $\Delta t=1$ day, respecting the CFL stability condition of the method (see^^>\cite{PTbook}). 

\begin{figure}[!p]
\centering
\includegraphics[width=0.32\textwidth]{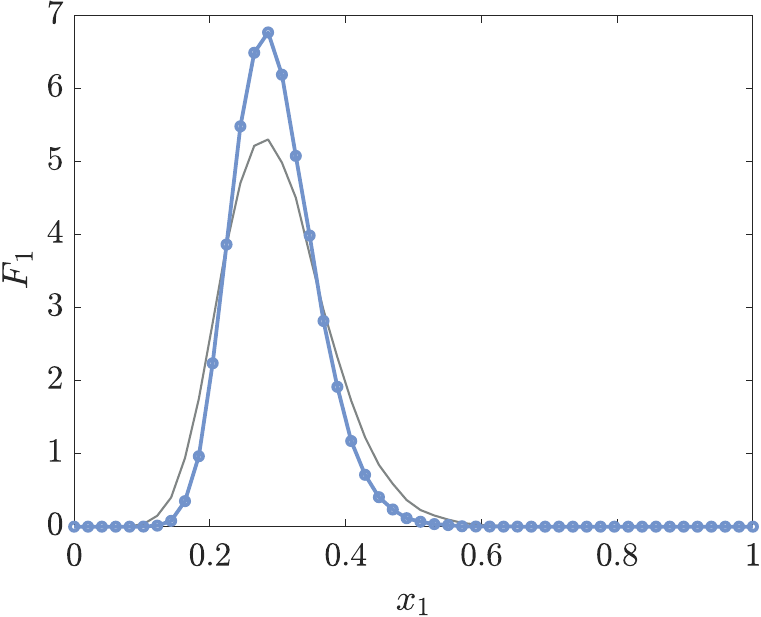}
\includegraphics[width=0.32\textwidth]{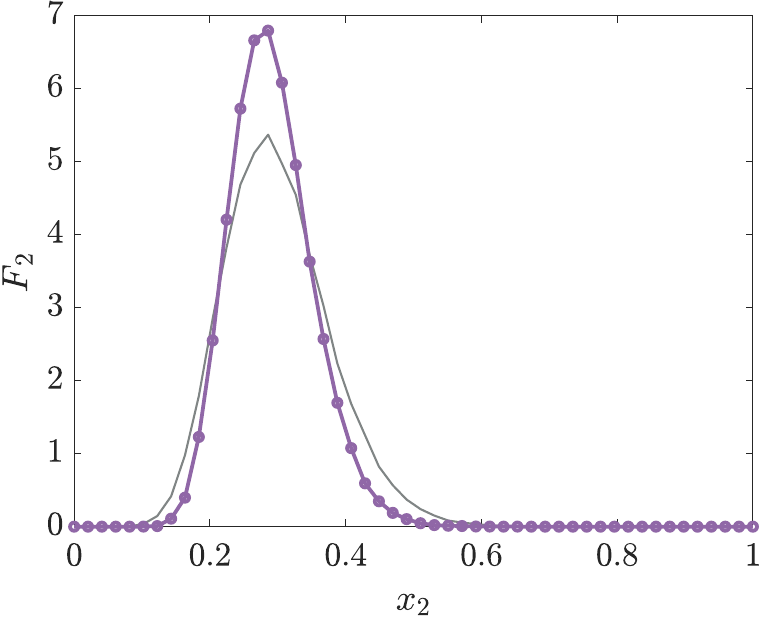}
\includegraphics[width=0.32\textwidth]{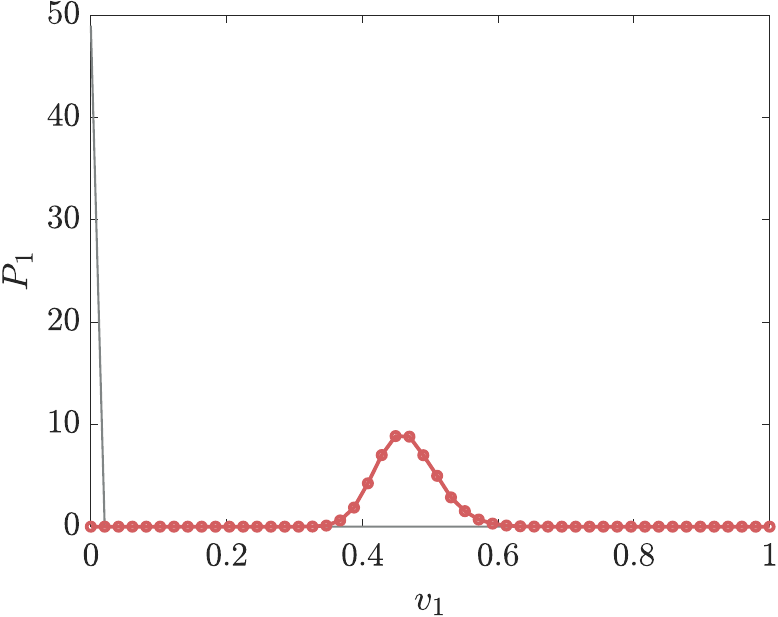}
\includegraphics[width=0.32\textwidth]{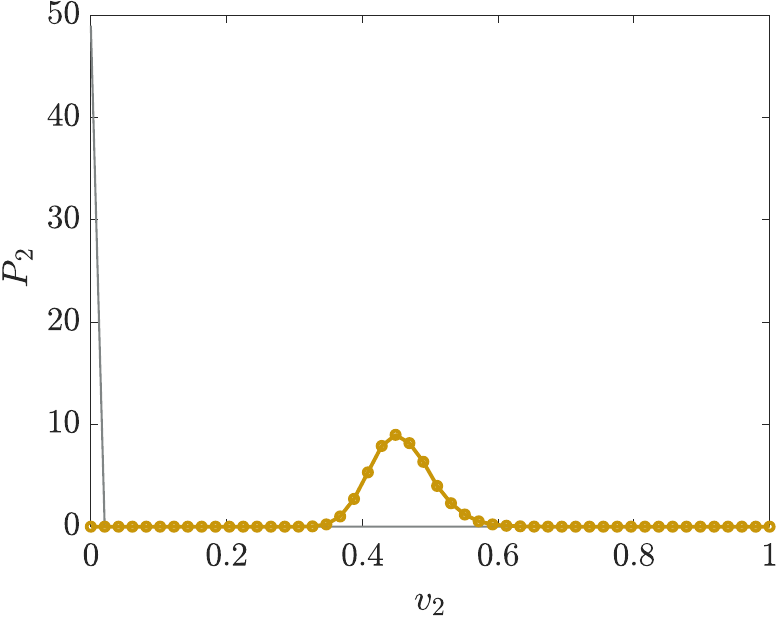}
\includegraphics[width=0.32\textwidth]{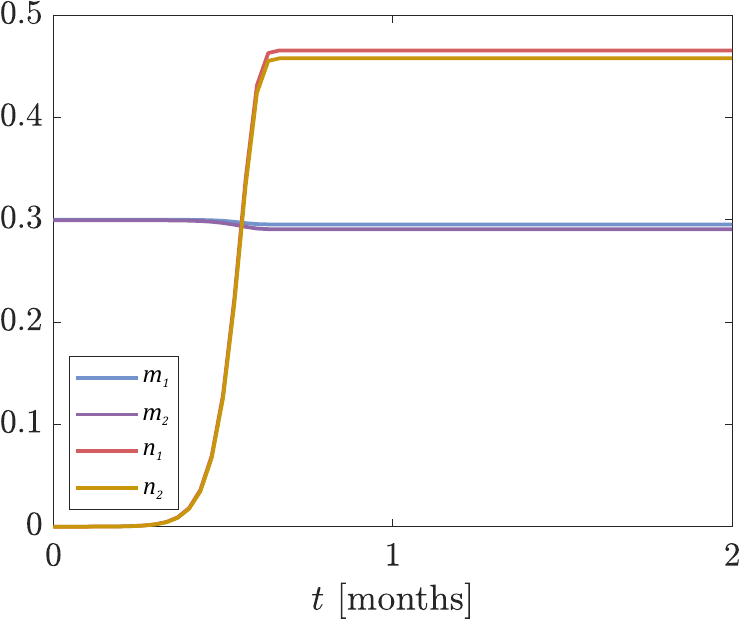}
\includegraphics[width=0.32\textwidth]{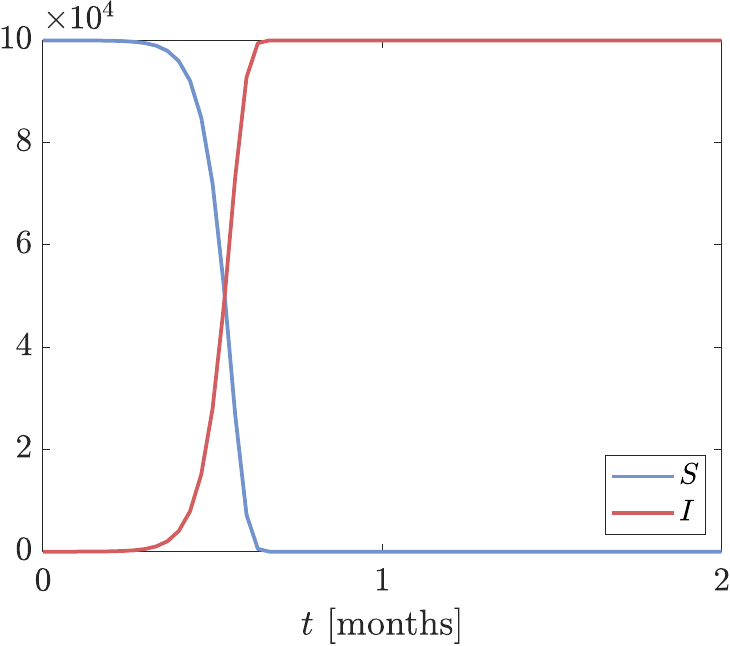}
\caption{Test 1(a), SI-type. Form left to right, top to bottom: \rev{distribution of resistance to the disease ($F_1$), sociality ($F_2$), contagiousness of the virus ($P_1$) and severity of the virus ($P_2$)} in the whole population at the final time $T = 60$ days (in colours) and at the initial time (in grey); evolution in time of the mean value of each state variable, and of $S$ and $I$ densities.}
\label{fig:test1_SI_kappa1}
\end{figure}
\begin{figure}[!p]
\centering
\includegraphics[width=0.32\textwidth]{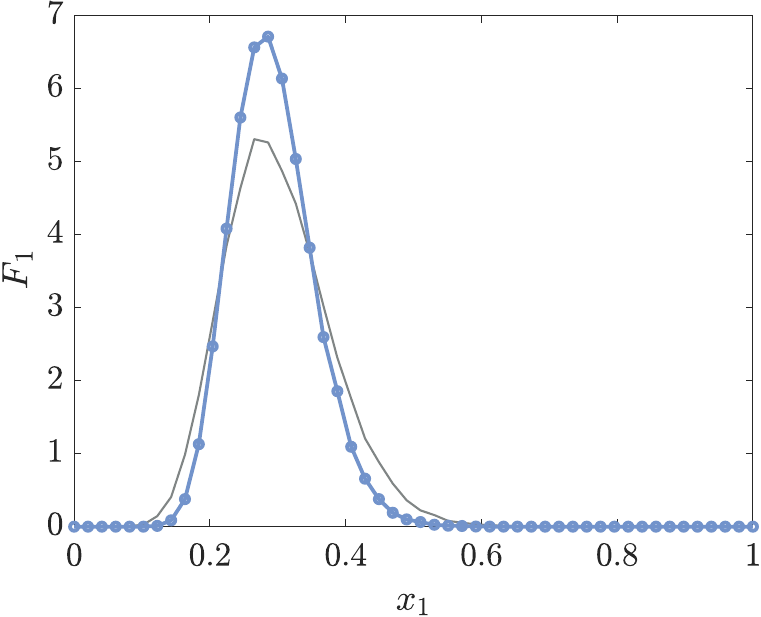}
\includegraphics[width=0.32\textwidth]{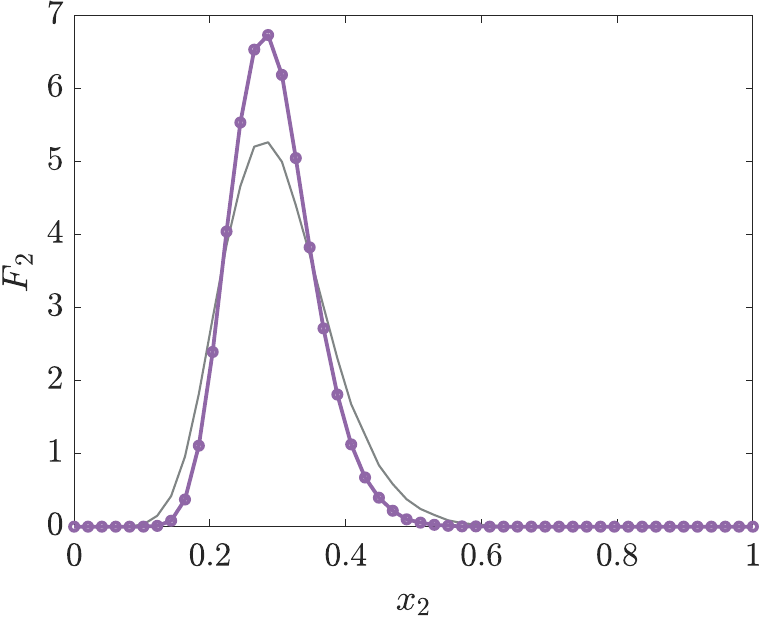}
\includegraphics[width=0.32\textwidth]{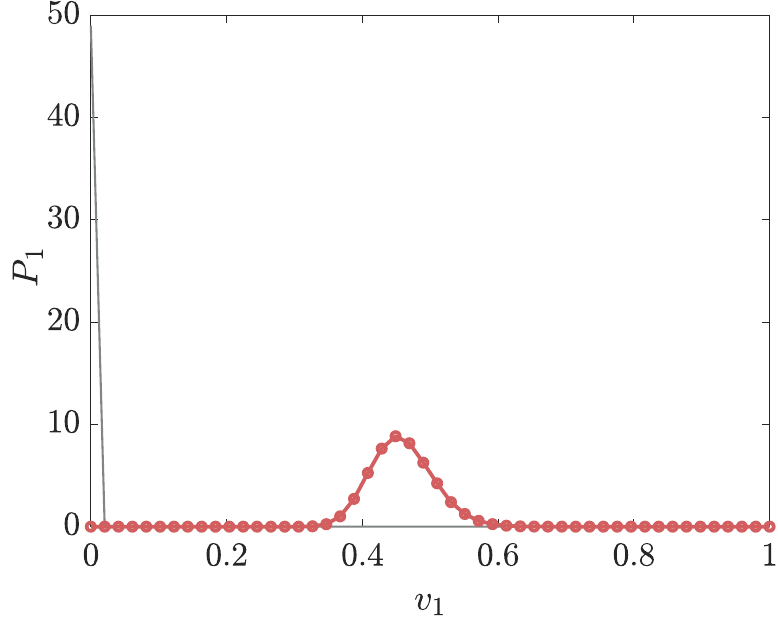}
\includegraphics[width=0.32\textwidth]{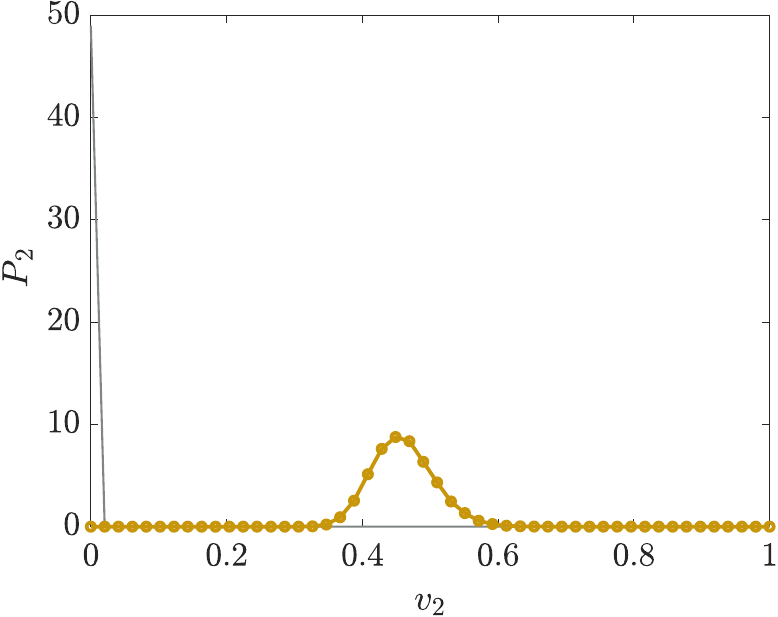}
\includegraphics[width=0.32\textwidth]{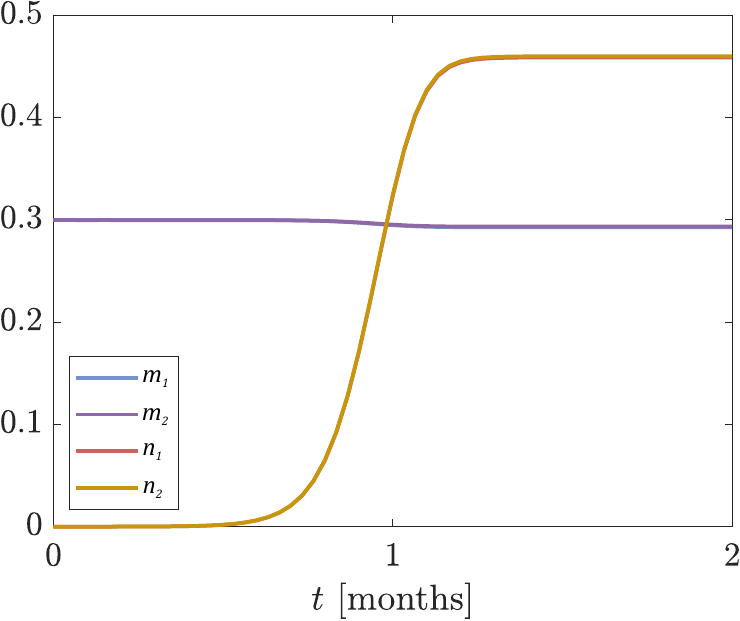}
\includegraphics[width=0.32\textwidth]{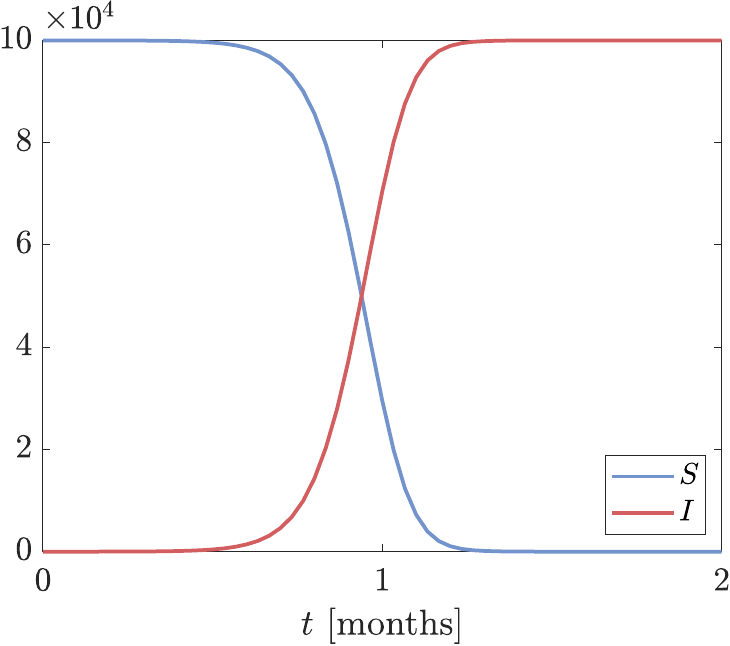}
\caption{Test 1(b), SI-type. Form left to right, top to bottom: \rev{distribution of resistance to the disease ($F_1$), sociality ($F_2$), contagiousness of the virus ($P_1$) and severity of the virus ($P_2$)} in the whole population at the final time $T = 60$ days (in colours) and at the initial time (in grey); evolution in time of the mean value of each state variable, and of $S$ and $I$ densities.}
\label{fig:test1_SI_kappa0.5}
\end{figure}
\begin{figure}[!p]
\centering
\includegraphics[width=0.32\textwidth]{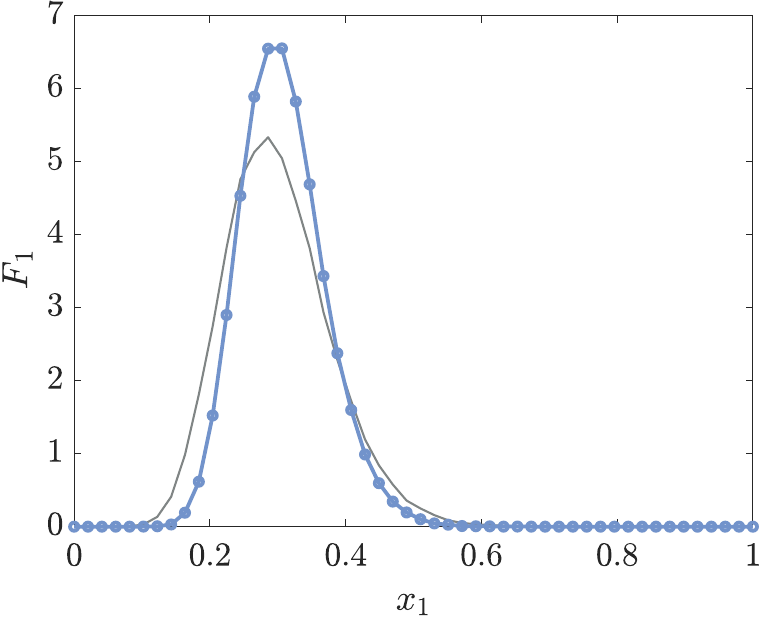}
\includegraphics[width=0.32\textwidth]{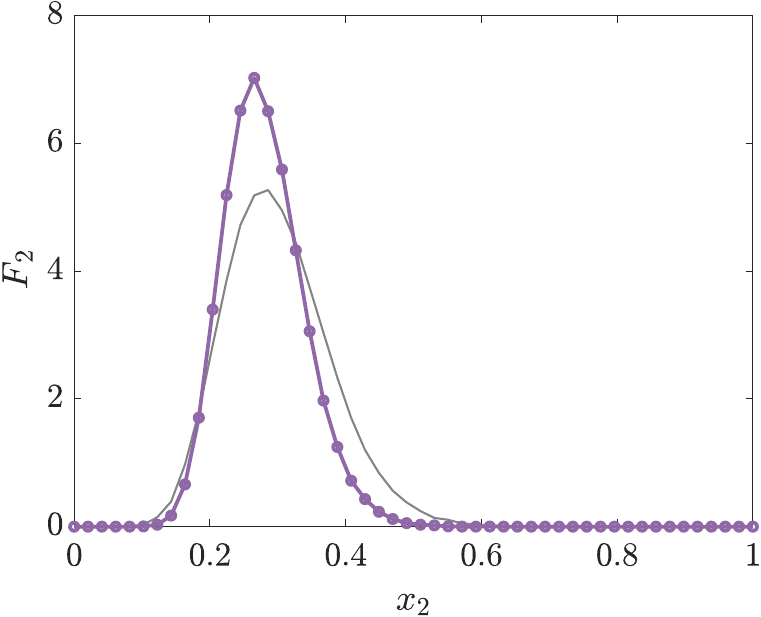}
\includegraphics[width=0.32\textwidth]{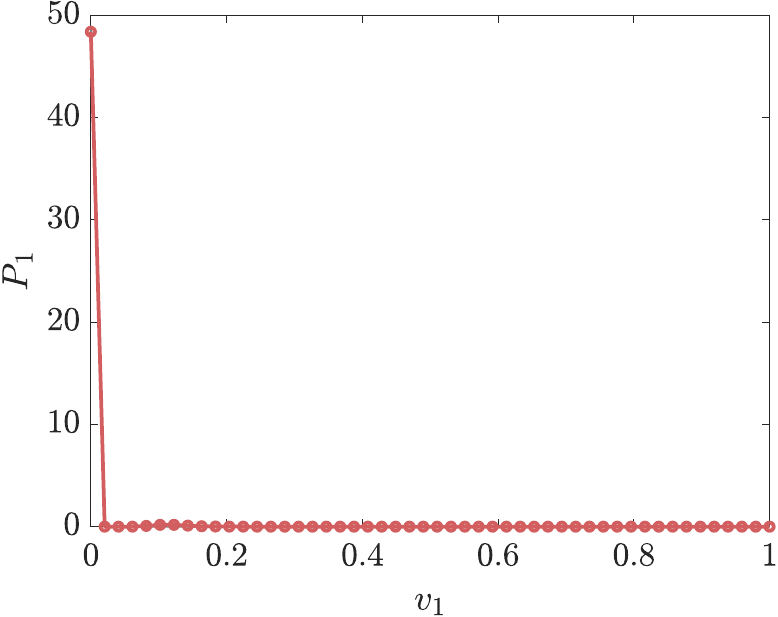}
\includegraphics[width=0.32\textwidth]{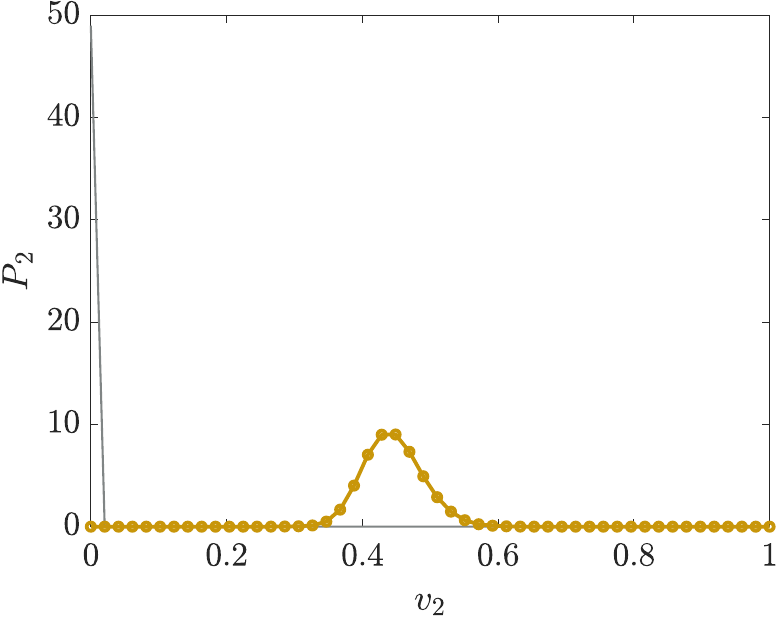}
\includegraphics[width=0.32\textwidth]{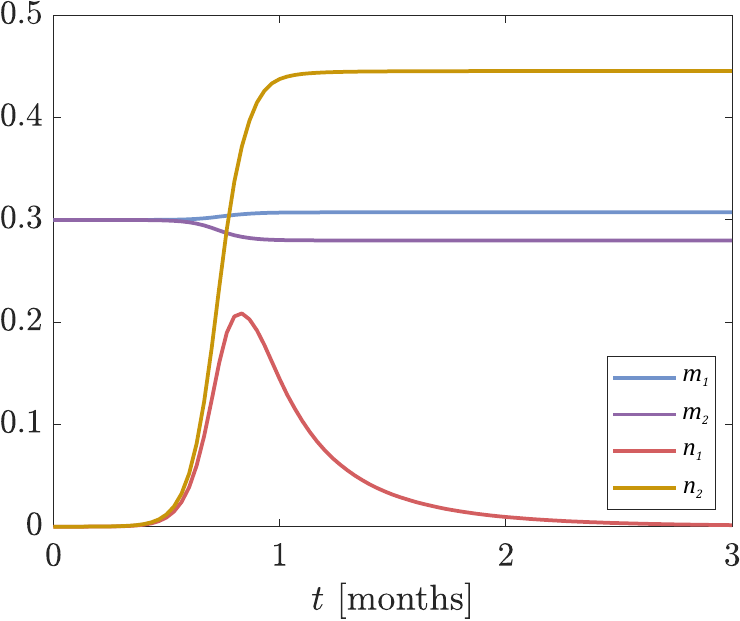}
\includegraphics[width=0.32\textwidth]{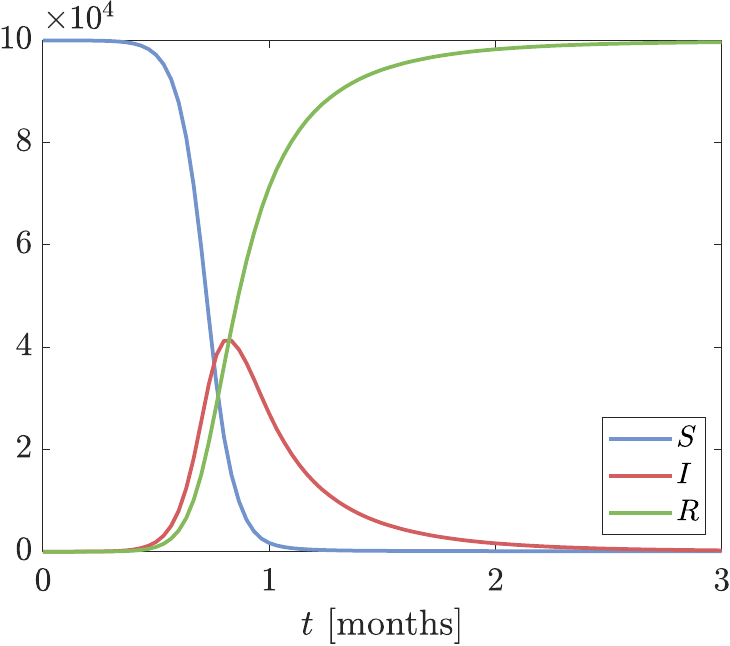}
\caption{Test 1(a), SIR-type. Form left to right, top to bottom: \rev{distribution of resistance to the disease ($F_1$), sociality ($F_2$), contagiousness of the virus ($P_1$) and severity of the virus ($P_2$)} in the whole population at the final time $T = 90$ days (in colours) and at the initial time (in grey); evolution in time of the mean value of each state variable, and of $S$, $I$ and $R$ densities.}
\label{fig:test1_SIR_kappa1}
\end{figure}
\begin{figure}[!p]
\centering
\includegraphics[width=0.32\textwidth]{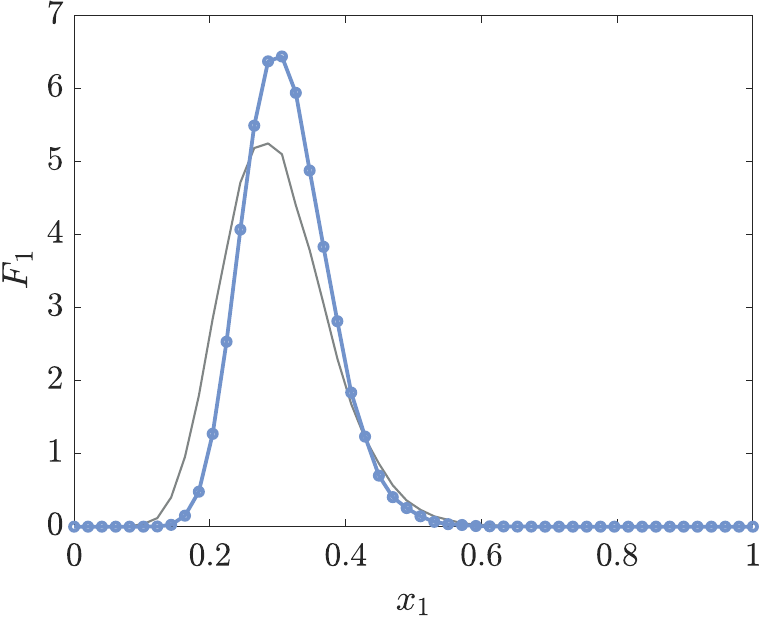}
\includegraphics[width=0.32\textwidth]{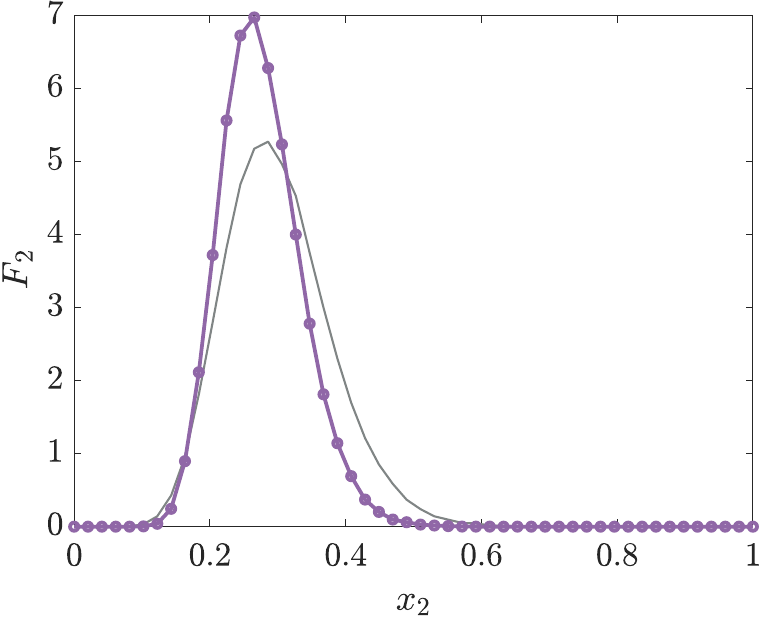}
\includegraphics[width=0.32\textwidth]{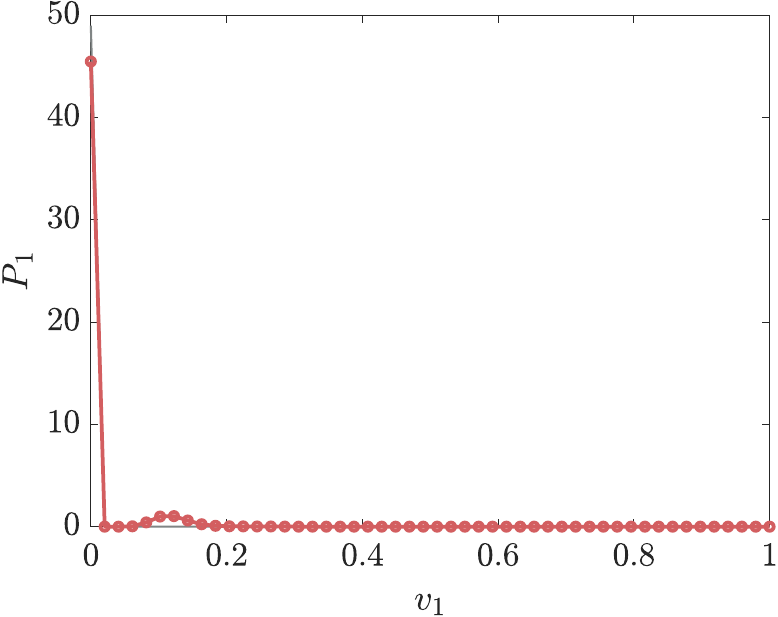}
\includegraphics[width=0.32\textwidth]{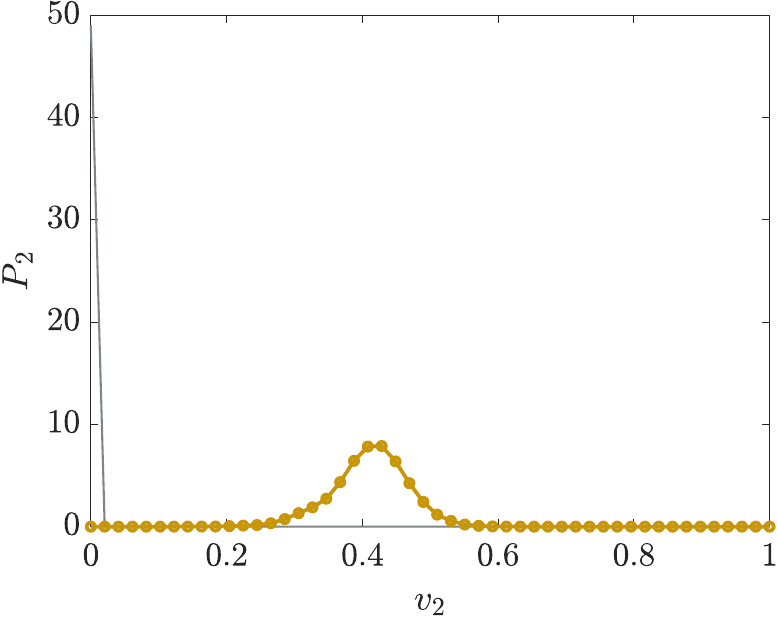}
\includegraphics[width=0.32\textwidth]{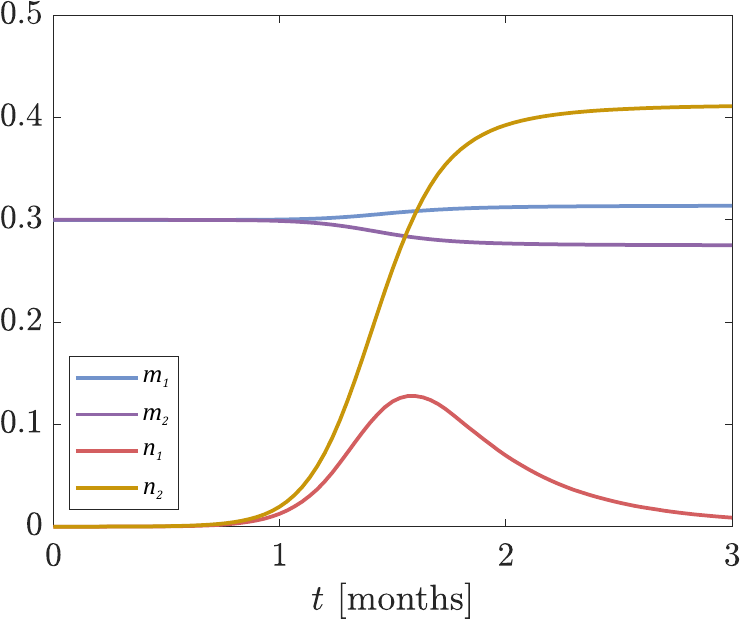}
\includegraphics[width=0.32\textwidth]{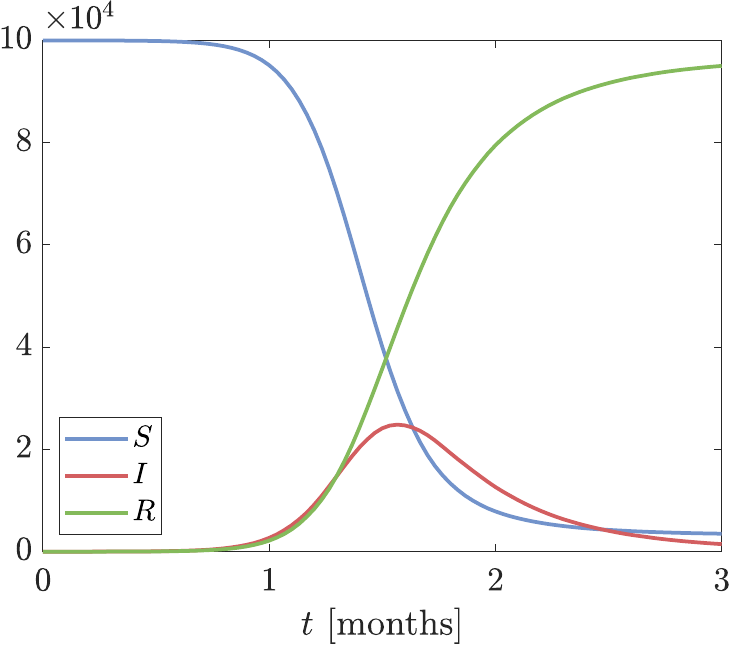}
\caption{Test 1(b), SIR-type. Form left to right, top to bottom: \rev{distribution of resistance to the disease ($F_1$), sociality ($F_2$), contagiousness of the virus ($P_1$) and severity of the virus ($P_2$)} in the whole population at the final time $T = 90$ days (in colours) and at the initial time (in grey); evolution in time of the mean value of each state variable, and of $S$, $I$ and $R$ densities.}
\label{fig:test1_SIR_kappa0.5}
\end{figure}
\begin{figure}[!t]
\centering
\includegraphics[width=0.32\textwidth]{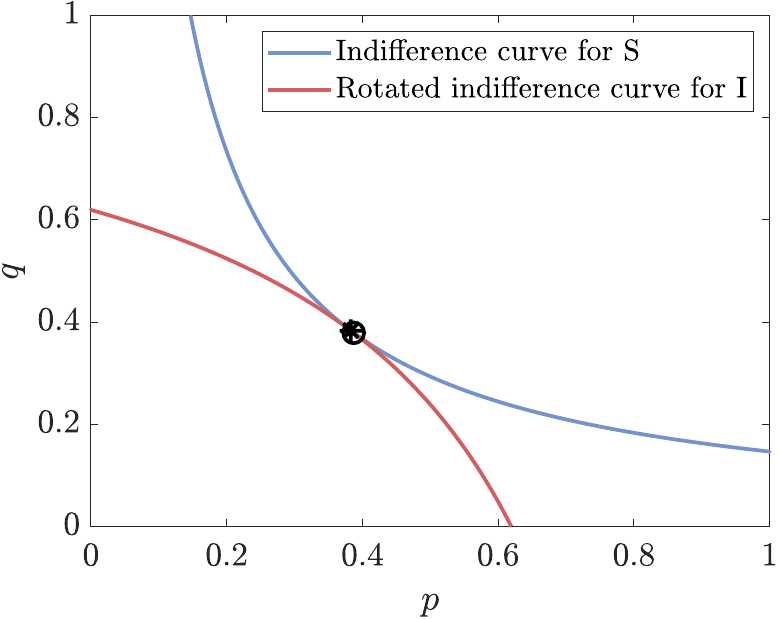}
\includegraphics[width=0.32\textwidth]{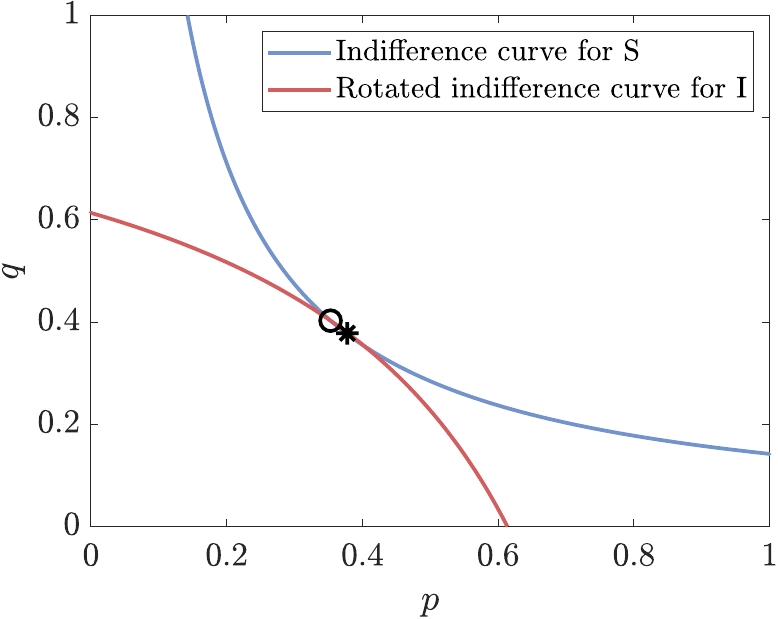}
\includegraphics[width=0.32\textwidth]{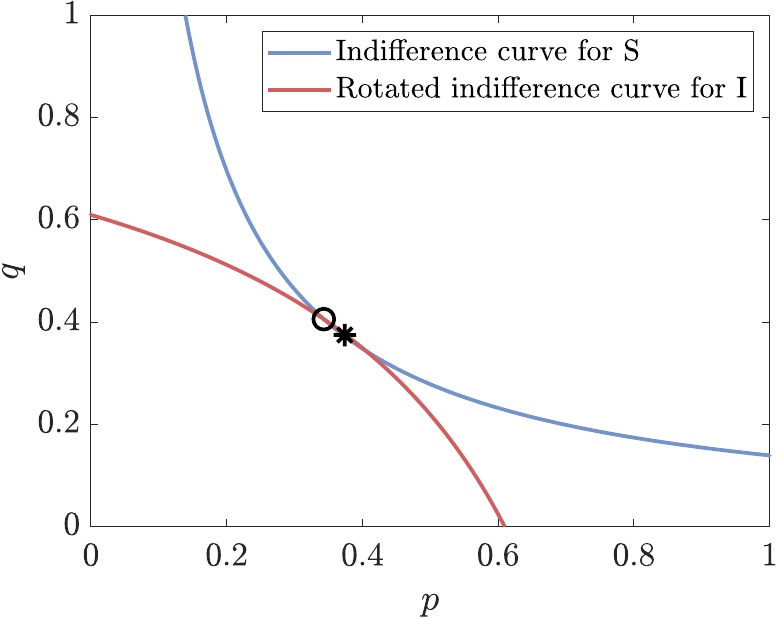}
\caption{Test 1(b), SIR-type. Infectious dynamics in the Edgeworth box at $t=1$, $t=45$, $t=90$ days (from left to right) averaged for all interactions that occurred at that time instant between the S-I pairs. The circle indicates the state pre-interaction, while the star depicts the state post-interaction.}
\label{fig:test1_SIR_kappa0.5_CDfunction}
\end{figure}

For each combination of cases, the average values resulting from 5 stochastic runs are shown in Figures \ref{fig:test1_SI_kappa1}-\ref{fig:test1_SIR_kappa0.5}.
It is immediate to observe, first of all, that the proposed methodology is able to recover the standard SI and SIR-type evolutions of the infectious disease by looking at the bottom-right plot of each Figure. In addition, it is possible to see what is the effect of the different choice of interaction kernel. Indeed, the halving of $\kappa$ leads to a significant slowdown in the spread of the virus, as expected \rev{(compare Fig. \ref{fig:test1_SI_kappa1} with \ref{fig:test1_SI_kappa0.5} and Fig. \ref{fig:test1_SIR_kappa1} with \ref{fig:test1_SIR_kappa0.5})}. The halving of $\kappa$, in fact, implies that there is only a 50\% chance (and no longer a 100\% chance) that a susceptible individual will become infected when it comes in contact with the infected agent. Finally, in the same figures it is possible to appreciate the variation in the marginal distributions of the state variables as well as the evolution of their mean value over time, noting that even when choosing $\alpha_S = \beta_S = 0.5$, the SIR dynamics leads to a different final mean for $x_1$ and $x_2$, favouring $x_1$ (survival rate to the disease).

Finally, in Figure \ref{fig:test1_SIR_kappa0.5_CDfunction} we present the dynamics of Test 1b, with SIR compartments, from the perspective of the Edgeworth box at 3 different time steps ($t=1$, $t=45$, $t=90$ days) averaged for all interactions that occurred at that time instant between the S-I pairs. We can observe that in all instances we are almost at the Pareto optimal.

\paragraph{Test 2.}
In the second numerical test, we stick directly to the SIR compartmentalization and consider a population of $N=10^4$ individuals among which 1\% are initially infected. We then assign different preference coefficients to the $S$ and $I$ sub-populations, choosing $\alpha_S = 0.7$, $\alpha_I = 0.4$ (hence, $\beta_S= 0.3$, $\beta_I = 0.6$). As previously done, we characterize both the uniformly distributed random variables $\mu$ and $\tilde\mu$ with mean $\lambda =0.49$ and variance $0.01$. Moreover, we assign an initial gamma distribution with shape and scale parameter respectively equal to 3 and 1, rescaled by a factor of 0.1, to both the socio-physical conditions $\ax$, and we attribute a random value of viral impact states $\av$ to the initially infected agents following a Gaussian distribution with mean 0.4 and variance 0.01.
In contrast with the previous test, in this one we consider a non-constant interaction kernel, \rev{defined as in \eqref{kappa}, setting} $\theta = 2$ and $\delta = \eta = 1$. The recovery rate is fixed to be $\gamma(\ax,\av) = 14^{-1}\, v_2/x_1$ days$^{-1}$.
We run the simulation up to the final time $T = 180$ days with $\Delta t=1/4$ of a day. In practice, with this choice, a single agent is assumed to interact 4 times with another random individual over the course of a day. 

\begin{figure}[!t]
\centering
\includegraphics[width=0.32\textwidth]{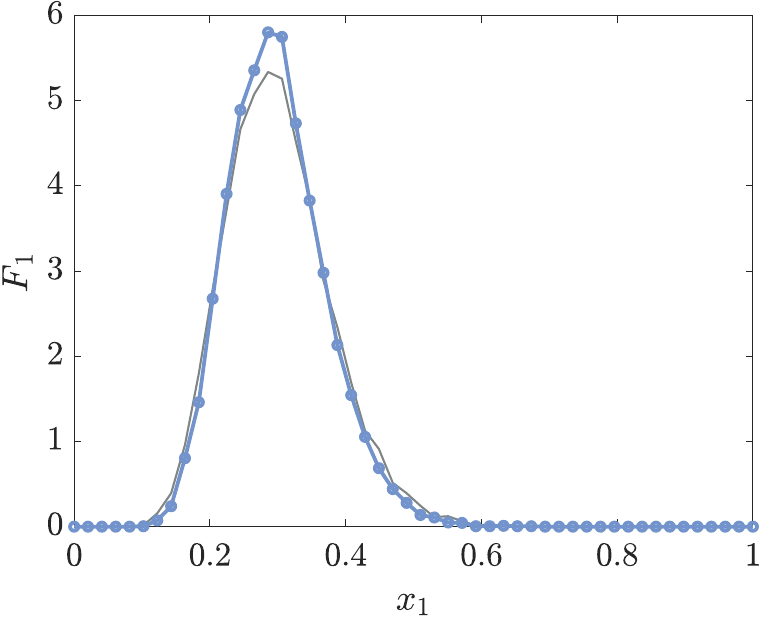}
\includegraphics[width=0.32\textwidth]{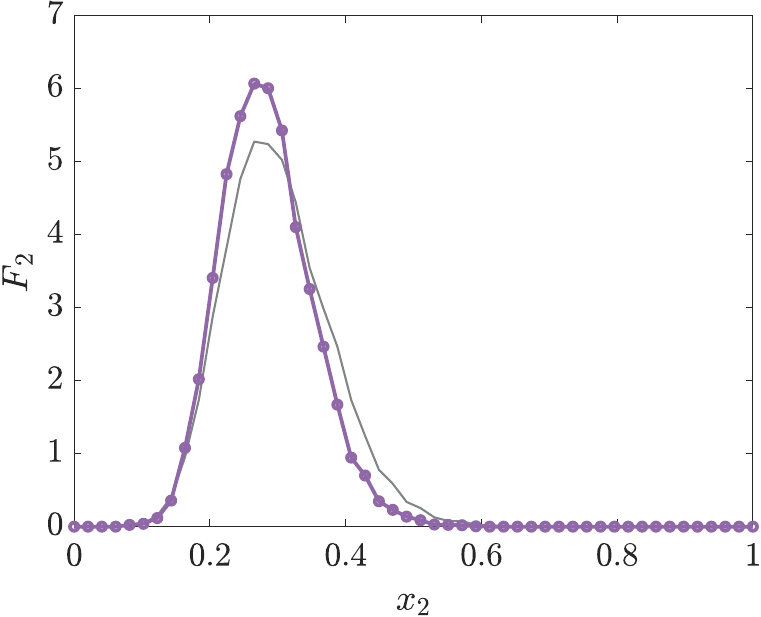}
\includegraphics[width=0.32\textwidth]{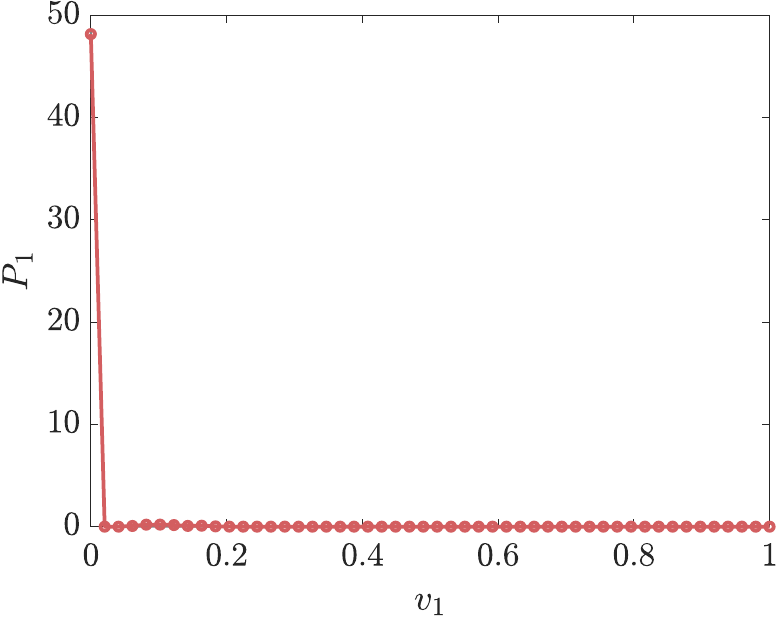}
\includegraphics[width=0.32\textwidth]{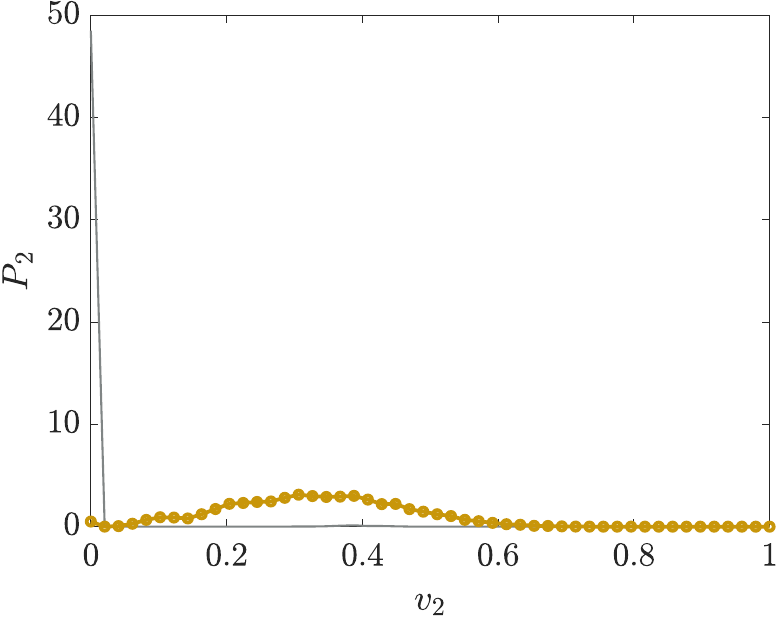}
\includegraphics[width=0.32\textwidth]{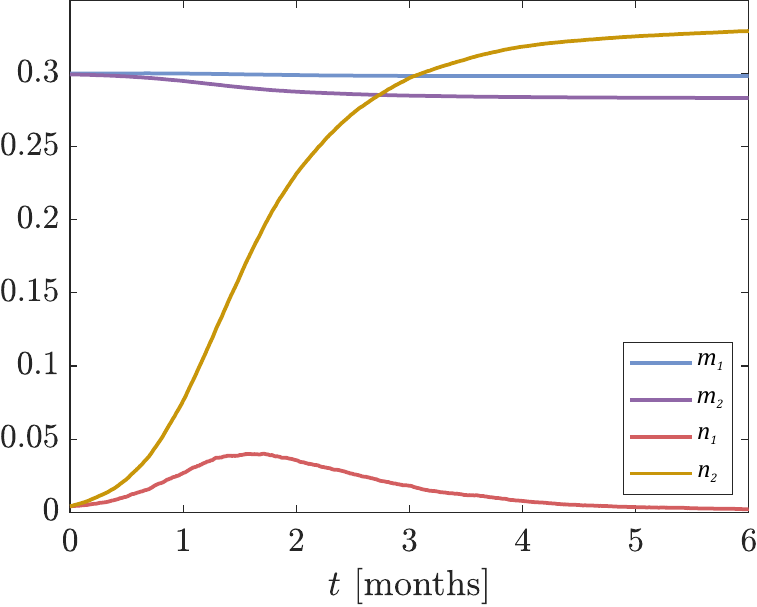}
\includegraphics[width=0.32\textwidth]{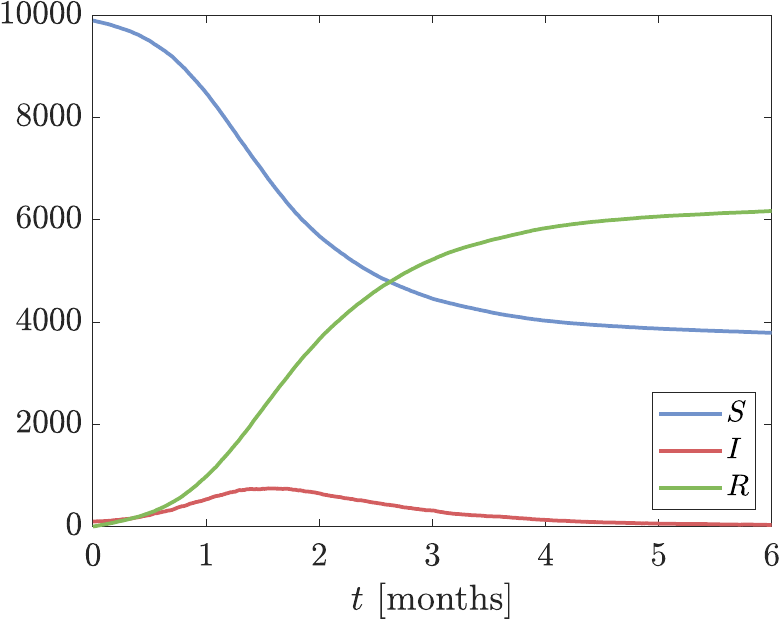}
\caption{Test 2. Form left to right, top to bottom: \rev{distribution of resistance to the disease ($F_1$), sociality ($F_2$), contagiousness of the virus ($P_1$) and severity of the virus ($P_2$)} in the whole population at the final time $T = 180$ days (in colours) and at the initial time (in grey); evolution in time of the mean value of each state variable, and of $S$, $I$ and $R$ densities.}
\label{fig:test2}
\end{figure}

We present the average values resulting from 5 stochastic runs in Figure \ref{fig:test2}. Here it is possible to observe that the chosen interaction kernel leads to a consistent containment of the spread of the virus with respect to the case of Test 1, making the evolution of the dynamics much less predictable. Also the choice of different preference coefficients $\alpha$ and $\beta$ enriches the dynamics and leads to a necessary reduction of the $\lambda$ value, due to the restriction \eqref{newb2} given by the Cobb-Douglas utility function, playing a significant role in the infectious process.

\paragraph{Test 3.}
In the last test, we consider again a population of $N=10^4$ individuals among which 1\% are initially infected, but this time we explore the effects of preference coefficients $\alpha$ and $\beta$ that are variable in time, simulating a sort of seasonal trends of these parameters. Thus, we define
\[ \alpha_S(t) = \beta_I(t) = 0.5 + 0.4 \cos\left(\frac{2\pi t}{T}\right) \qquad \alpha_I(t) = \beta_S(t) = 0.5 - 0.4 \cos\left(\frac{2\pi t}{T}\right)\,,\]
with $T=1$ year. With this choice, we are hypothesizing that during the spring-summer season individuals will tend to prefer sociality over disease resistance, while, in the same time frame, the virus will tend to favour contagiousness over severity. The opposite preferences will apply for the autumn-winter season. As a consequence, also the coefficient $\lambda \coloneqq \lambda(t)$, varying in time while respecting restriction \eqref{newb2}.

\begin{figure}[!p]
\centering
\includegraphics[width=0.32\textwidth]{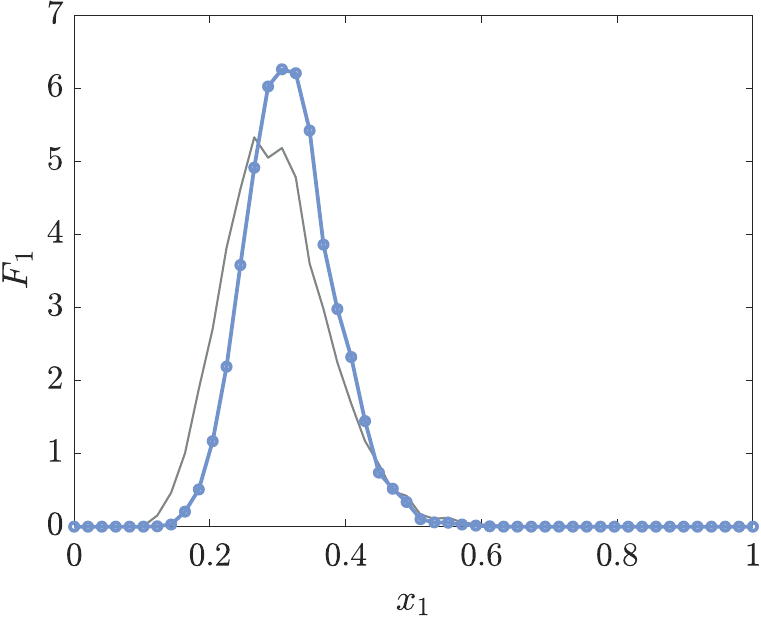}
\includegraphics[width=0.32\textwidth]{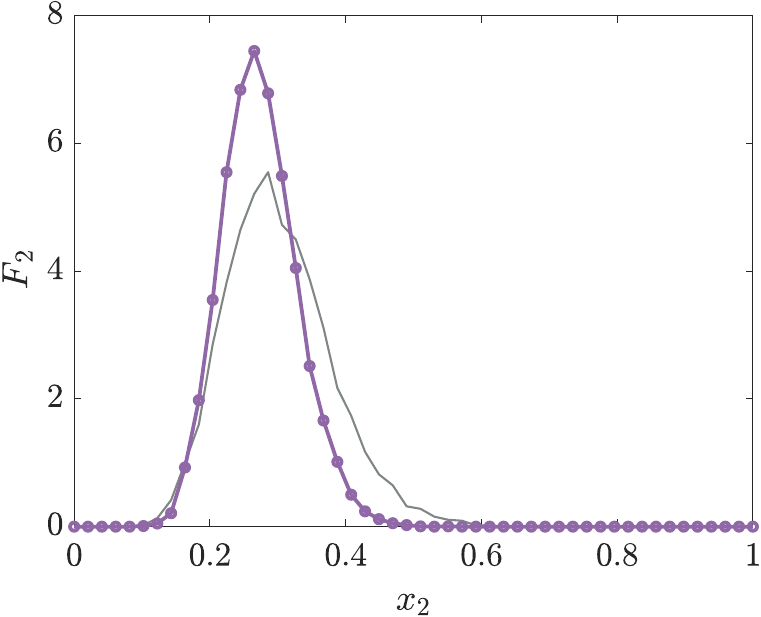}
\includegraphics[width=0.32\textwidth]{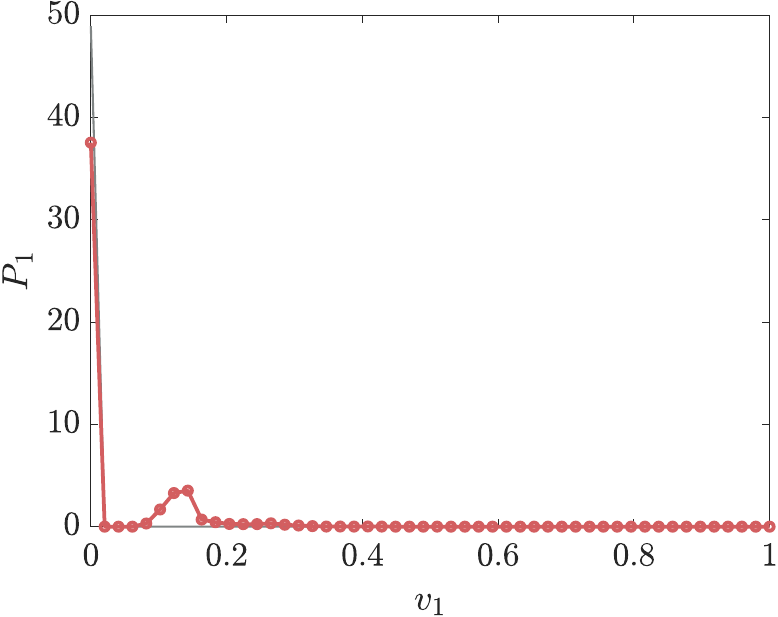}
\includegraphics[width=0.32\textwidth]{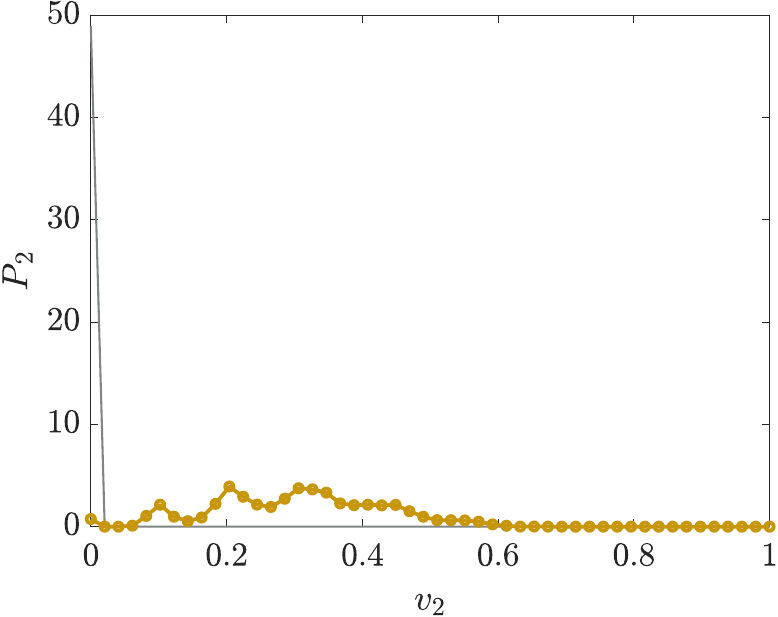}
\includegraphics[width=0.32\textwidth]{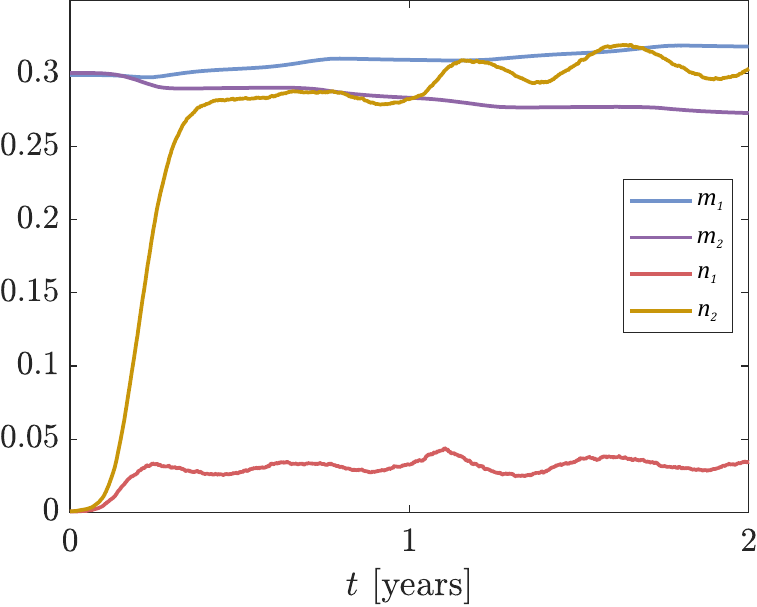}
\includegraphics[width=0.32\textwidth]{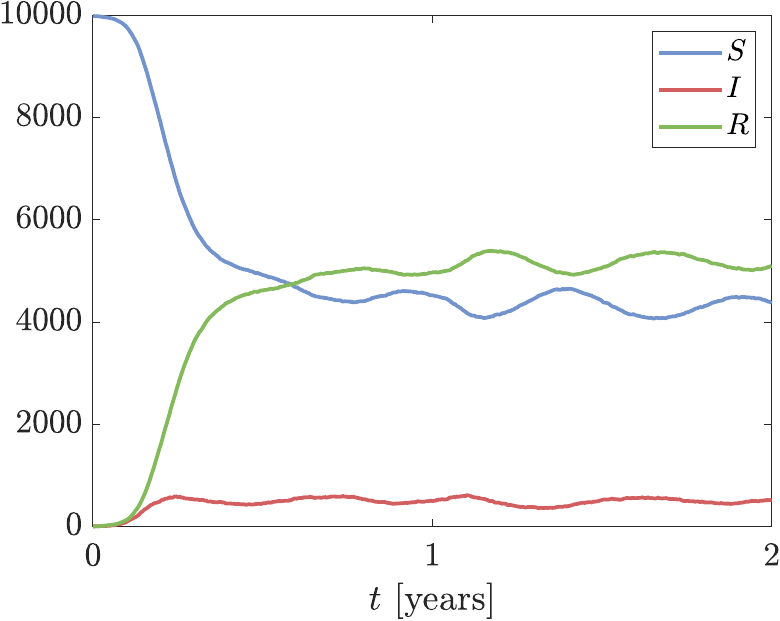}
\caption{Test 3(a). Form left to right, top to bottom: \rev{distribution of resistance to the disease ($F_1$), sociality ($F_2$), contagiousness of the virus ($P_1$) and severity of the virus ($P_2$)} in the whole population at the final time $2T = 2$ years (in colours) and at the initial time (in grey); evolution in time of the mean value of each state variable, and of $S$, $I$ and $R$ densities.}
\label{fig:test3a}
\end{figure}
\begin{figure}[!p]
\centering
\includegraphics[width=0.32\textwidth]{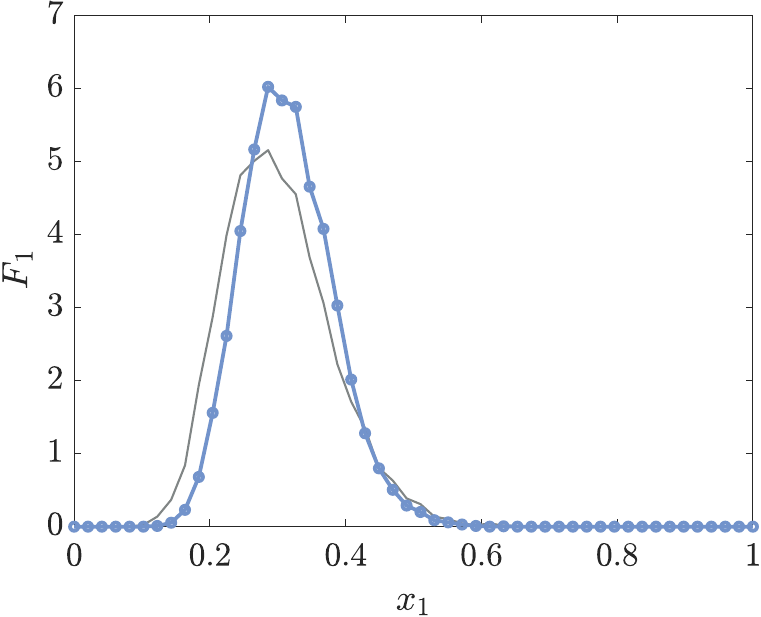}
\includegraphics[width=0.32\textwidth]{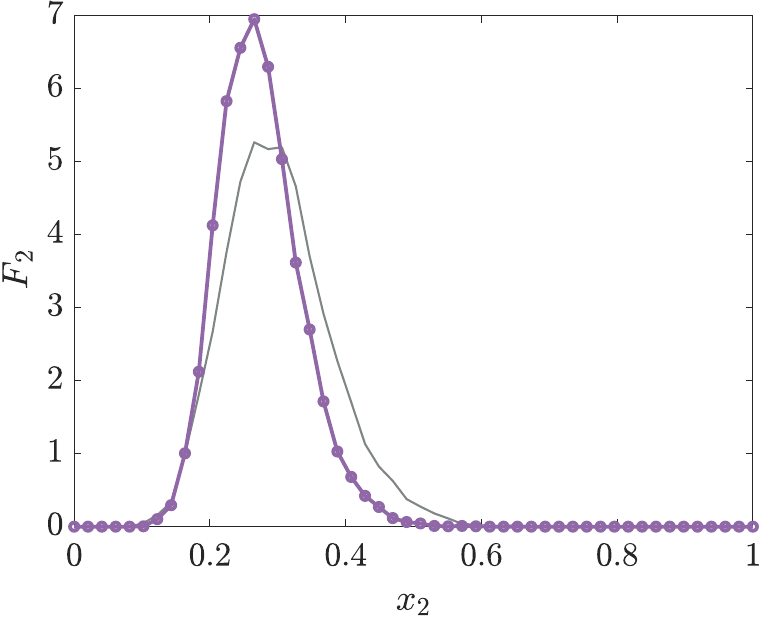}
\includegraphics[width=0.32\textwidth]{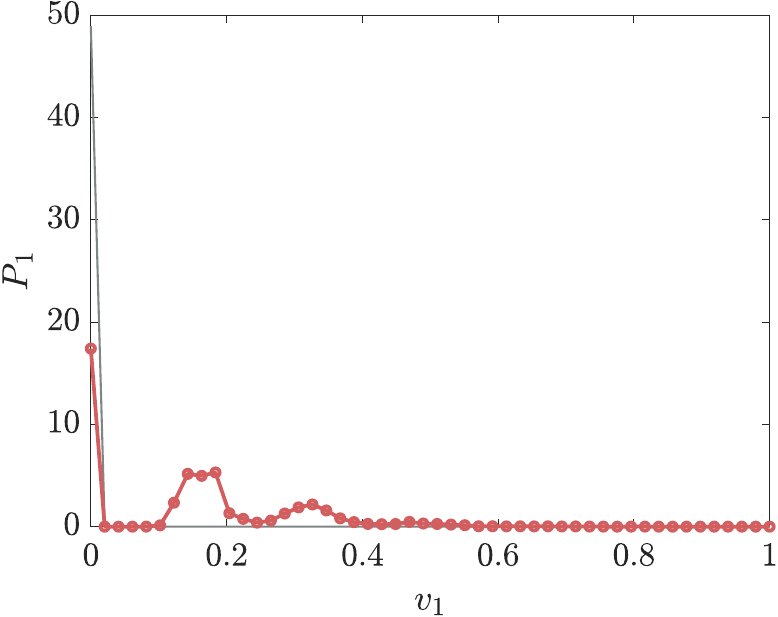}
\includegraphics[width=0.32\textwidth]{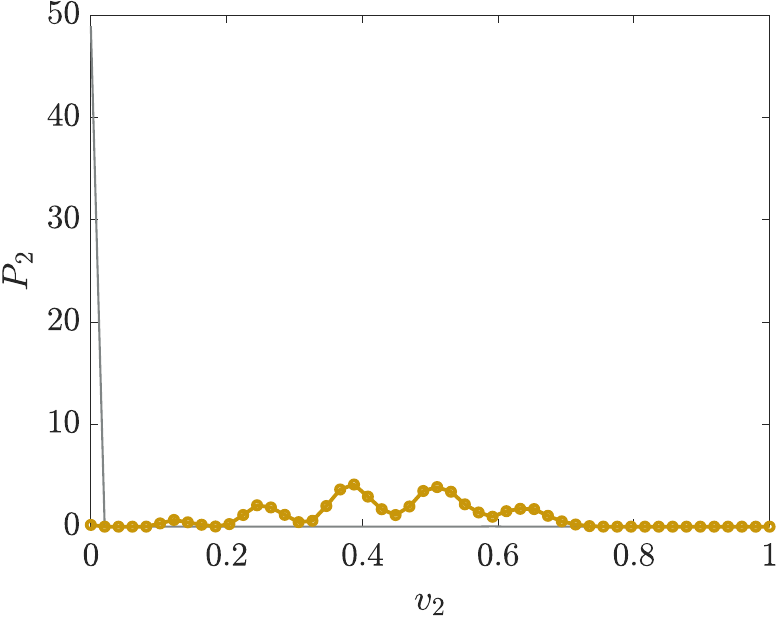}
\includegraphics[width=0.32\textwidth]{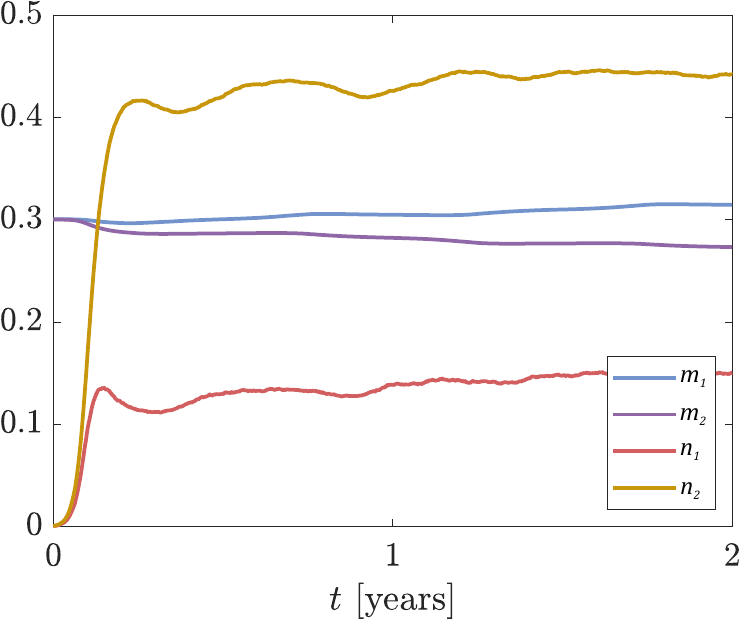}
\includegraphics[width=0.32\textwidth]{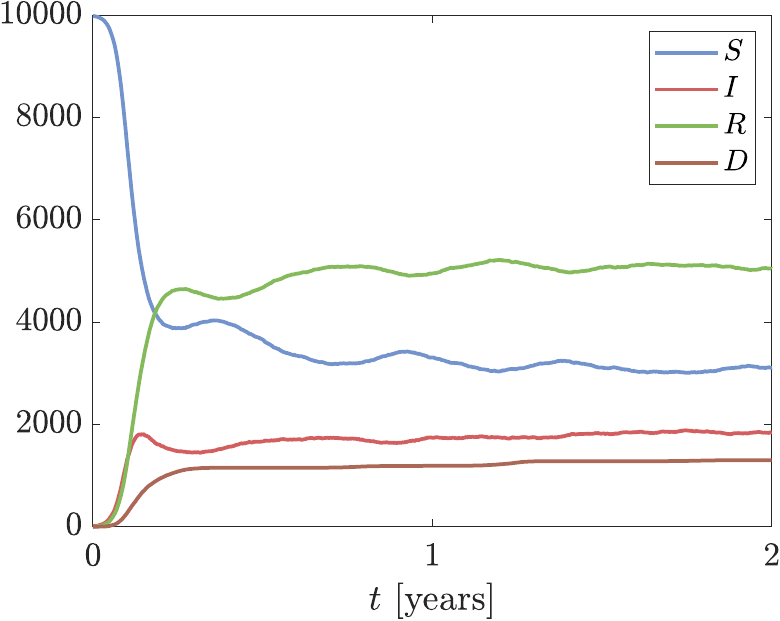}
\caption{Test 3(b). Form left to right, top to bottom: \rev{distribution of resistance to the disease ($F_1$), sociality ($F_2$), contagiousness of the virus ($P_1$) and severity of the virus ($P_2$)} in the whole population at the final time $2T = 2$ years (in colours) and at the initial time (in grey); evolution in time of the mean value of each state variable, and of $S$, $I$, $R$ and $D$ densities.}
\label{fig:test3b}
\end{figure}
\begin{figure}[!t]
\centering
\includegraphics[width=0.48\textwidth]{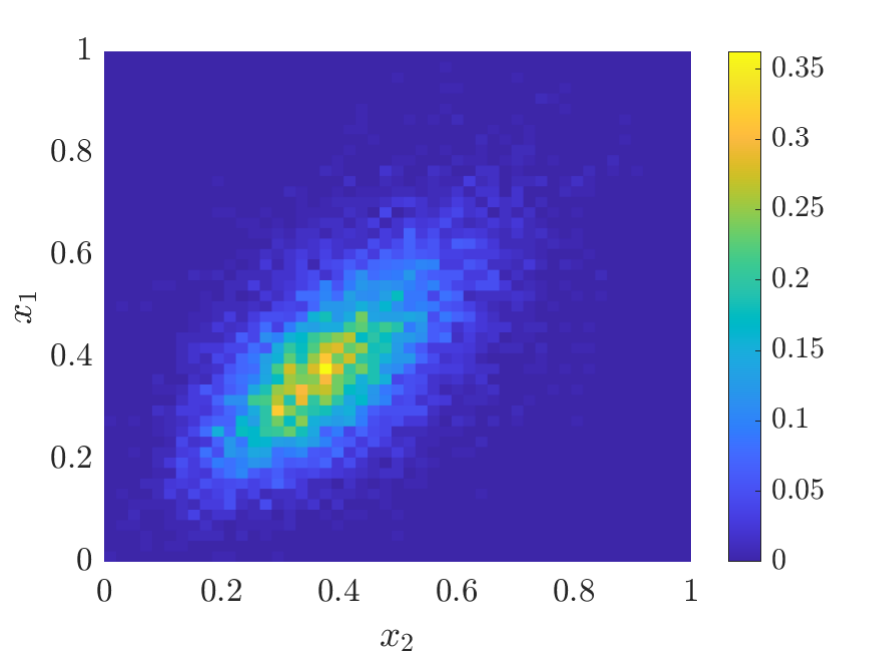}
\includegraphics[width=0.48\textwidth]{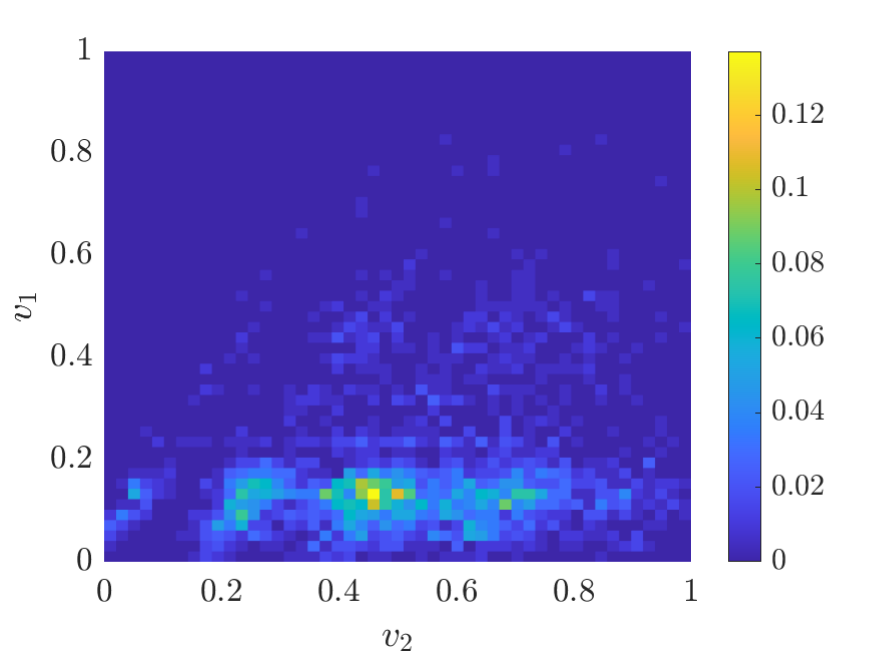}
\includegraphics[width=0.48\textwidth]{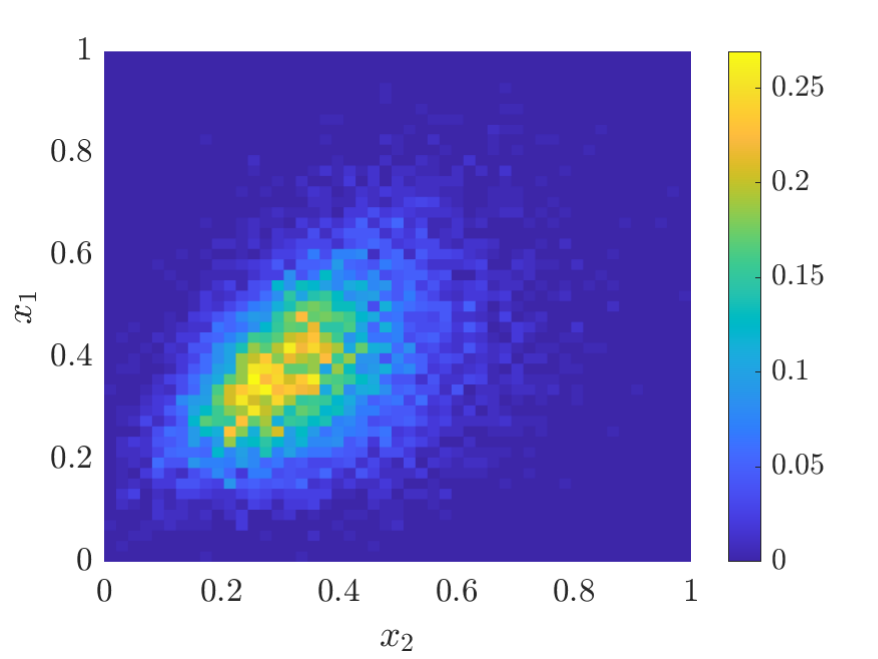}
\includegraphics[width=0.48\textwidth]{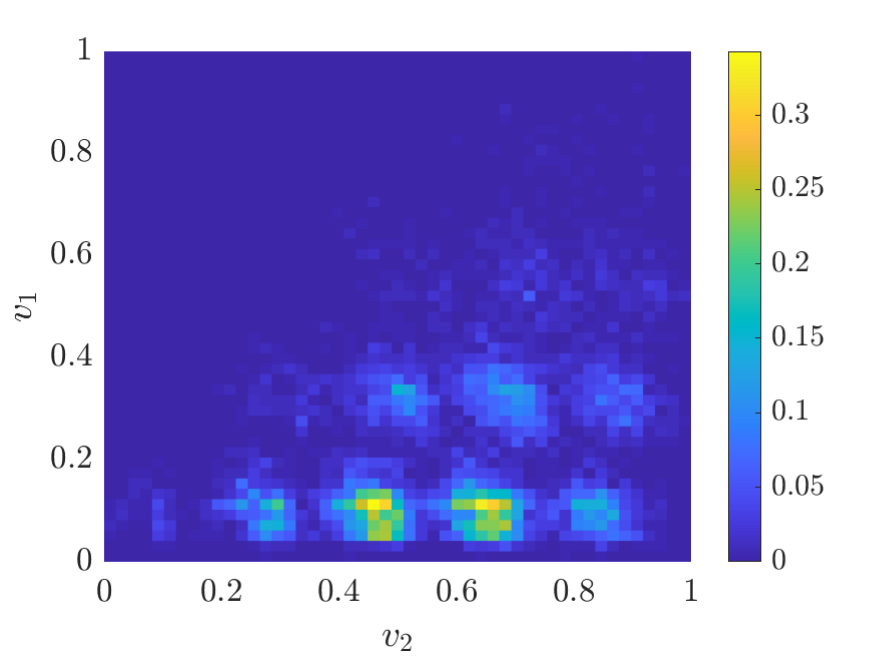}
\caption{Test 3. Bivariate distribution of socio-physical condition \rev{$F$} (left) and viral impact \rev{$P_I$} (right) at the final time in test case (a) (first row) and (b) (second row). The viral impact is presented taking into account only the contribution of infected agents.}
\label{fig:test3_Fxv}
\end{figure}

In addition, in this test we introduce the dynamics of loss of immunity, considering that a healed individual will lose immunity to the virus 2 months after the recovery. This dynamics is introduced, at the modelling level, in a manner entirely similar to that of healing. Thus, the loss of immunity rate is fixed to be $\sigma(\ax,\av) = 60^{-1}\, x_1/v_2$ days$^{-1}$. 

We simulate two different configurations, corresponding to case (a) and case (b).
In Test 3(a) the rest of the parameters, including the initial conditions, are left unchanged with respect to Test 2.
In Test 3(b) we simulate the spread of a more severe virus by changing the initial distribution of the variables $\av$, this time following a Gaussian with mean 0.6 (instead of 0.4), and we increase the mean time to recovery to 20 days (i.e., $\gamma(\ax,\av) = 20^{-1}\, v_2/x_1$ days$^{-1}$). 
Furthermore, in Test 3(b) we consider the process of death as discussed in Section \ref{sec:compartments}, by setting an upper limit of severity $\hat v_2 = 0.75$ beyond which the individual is considered deceased. This approach essentially allows us to introduce an additional compartment $D$ of deceased individuals into the model.
Both simulations are performed up to $2T = 2$ years, again fixing $\Delta t=1/4$ of a day.

We present numerical results obtained averaging 5 stochastic runs in Figures \ref{fig:test3a}-\ref{fig:test3b}. From the last two plots of both figures, the effect of the seasonal variation of the preference coefficients can be clearly appreciated, and we observe that the epidemic tends to reach an endemic state. The final distribution of the \rev{severity of the virus ($P_2$)} also appears much more varied in both the configurations. On the other hand, the distributions of the socio-physical variables \rev{($F_1$ and $F_2$)} seem to be less affected by the variability of $\alpha$ and $\beta$. To highlight this effect, in Figure \ref{fig:test3_Fxv} we show the bivariate distribution of the socio-physical condition of the whole population and of the viral impact related to infected agents at the final time of the simulations. Concerning only Test 3(b), in Figure \ref{fig:test3b} we can see that the death process appears significant only in the early stages of infectious disease spread and, instead, remains under control in the later stage of endemic equilibrium. 

\section{Conclusion}
In this paper, we explore the realm of irreversibility in biological processes, employing a kinetic modelling framework grounded in a utility theory approach. \rev{Our primary focus has been on understanding the dynamics of multi-agent systems in the context of infectious diseases. The proposed modeling is based on a new characterization of the transmission of infection that takes into account both individual-specific aspects, such as sociality and viral load, and intrinsic characteristics related to the severity of the virus.} This approach, drawing inspiration from microeconomics and the renowned Edgeworth box, complements existing models, such as those employing game theory^^>\cite{Bohl, Cas}, \rev{to characterize the dynamics of virus spread.} 

\rev{Our exploration of irreversibility, referring to phenomena where changes within the system cannot be reversed, led us to introduce the concept of ``utility" as a driving force for interactions between agents and viruses. This innovative approach has allowed us to propose kinetic equations of Boltzmann-type, providing a dynamic description of the probability distributions in agent-virus systems undergoing binary interactions. Our methodology, complementing other approaches in the field, provides a valuable tool for gaining insights into the dynamics of infectious diseases and their transition from high-risk pandemic phenomena to low-risk endemic states.}
The numerical experiments presented, indeed, shed light on the mechanism steering these phenomena toward endemicity, specifically within the oscillatory nature observed in successive pandemic waves. 

In conclusion, this exploration, albeit preliminary, aims to lay the foundations for a new modelling approach to epidemiology based on utility functions and serves as a basis for further investigations into the complexity of infectious disease dynamics, \rev{with potential applications in public health and disease management.}


\section*{Acknowledgments} 
This work has been written within the activities of GNCS and GNFM groups of INdAM (Italian National Institute of High Mathematics). LP has been supported by the Royal Society under the Wolfson Fellowship ``Uncertainty quantification, data-driven simulations and learning of multiscale complex systems governed by PDEs". The partial support by ICSC -- Centro Nazionale di Ricerca in High Performance Computing, Big Data and Quantum Computing, funded by European Union -- NextGenerationEU and by MUR PRIN 2022 Project No. 2022KKJP4X ``Advanced numerical methods for time dependent parametric partial differential equations with applications" is also acknowledged. GB has also been partially funded by the call ``Bando Giovani anno 2023 per progetti di ricerca finanziati con il contributo 5x1000 anno 2021'' of the University of Ferrara,
and by the European Union -- NextGenerationEU, MUR PRIN 2022 PNRR Project No. P2022JC95T ``Data-driven discovery and control of multi-scale interacting artificial agent systems''.
\medskip


\end{document}